\documentclass[amsmath, amssymb, aps, reprint,  superscriptaddress]{revtex4-2}
\usepackage{siunitx}
\usepackage{graphicx}
\usepackage{dcolumn}
\usepackage{bm}
\usepackage{hyperref}
\usepackage{xfrac}
\usepackage{textgreek}
\usepackage{xspace}
\usepackage{nicefrac}

\hypersetup{hidelinks}

\def\mem{\kern-1pt\triangleright\kern-1pt}
\def\den{\rho}
\def\sup{\textsc{s}}

\newcommand{\one}{A}
\newcommand{\two}{B}

\newcommand{\methods}{Sec.\xspace}

\newcommand{\pump}{\SI{1.0}{\kilo\electronvolt}\xspace}
\newcommand{\probe}{\SI{1.2}{\kilo\electronvolt}\xspace}
 
\graphicspath{{./figures/}}

\newcommand{\affila}{Laboratory for Solid State Physics, ETH Zurich, 8093 Zürich, Switzerland}
\newcommand{\affilb}{Paul Scherrer Institut, 5232 Villigen, Switzerland}
\newcommand{\affilc}{Wilhelm Ostwald Institute for Physical and Theoretical Chemistry, Leipzig University, 04103 Leipzig, Germany}
\newcommand{\affild}{European XFEL, 22869 Schenefeld, Germany}
\newcommand{\affile}{Laboratory for Ultrafast X-ray Sciences, Institute of Chemical Sciences and Engineering, Ecole Polytechnique Federale de Lausanne (EPFL), 1015 Lausanne, Switzerland}
\newcommand{\affilf}{Quantum Center of Excellence for Diamond and Emergent Materials and Department of Physics, Indian Institute of Technology Madras, Chennai 600036, India}
\newcommand{\affilg}{Institute of Physics, University of Rostock, 18051 Rostock, Germany}
\newcommand{\affilh}{Department of Chemistry and Biotechnology, Darmstadt University of Applied Sciences, 64295 Darmstadt, Germany}
\newcommand{\affili}{EUt+ Institute of Nanomaterials \& Nanotechnologies EUTINN, European University of Technology, European Union}
\newcommand{\affilj}{PNSensor GmbH, 81739 Munich, Germany}
\newcommand{\affilk}{Institute of Physics, University of Kassel, 34132 Kassel, Germany}
\newcommand{\affill}{Institute for Optics and Atomic physics, Technical University Berlin, 10623 Berlin, Germany}
\newcommand{\affilm}{Max-Planck-Institut für Kernphysik, 69117 Heidelberg, Germany}
\newcommand{\affiln}{Institute for Electronics, ETH Zurich, 8049 Zurich, Switzerland}
\newcommand{\affilo}{Institute of Physics, University of Freiburg, 79104 Freiburg, Germany }
\newcommand{\affilp}{Department of Chemistry, Aarhus University, 8000 Aarhus C, Denmark}

\usepackage{pdfpages}
\makeatletter
\AtBeginDocument{\let\LS@rot\@undefined}
\makeatother

\newcommand{\newtextcolor}{black}

\newcommand{\newtext}[1]{\textcolor{\newtextcolor}{#1}}

\renewcommand{\vec}{\mathbf}

\begin{document}

\title{Dichography: Two-frame Ultrafast Imaging from a Single Diffraction Pattern}

\author{Linos \surname{Hecht}}\affiliation{\affila}
\author{Andre \surname{Al Haddad}}\affiliation{\affilb}
\author{Björn \surname{Bastian}}\affiliation{\affilc}
\author{Thomas M. \surname{Baumann}}\affiliation{\affild}
\author{Johan \surname{Bielecki}}\affiliation{\affild}
\author{Christoph \surname{Bostedt}}\affiliation{\affilb}\affiliation{\affile}
\author{Subhendu \surname{De}}\affiliation{\affilf}
\author{Alberto \surname{De Fanis}}\affiliation{\affild}
\author{Simon \surname{Dold}}\affiliation{\affild}
\author{Thomas \surname{Fennel}}\affiliation{\affilg}
\author{Fanny \surname{Goy}}\affiliation{\affila}
\author{Christina \surname{Graf}}\affiliation{\affilh}\affiliation{\affili}
\author{Robert \surname{Hartmann}}\affiliation{\affilj}
\author{Georg \surname{Jakobs}}\affiliation{\affila}
\author{Maximilian \surname{Joschko}}\affiliation{\affilh}\affiliation{\affili}
\author{Gregor \surname{Knopp}}\affiliation{\affilb}
\author{Katharina \surname{Kolatzki}}\affiliation{\affila}
\author{Sivarama \surname{Krishnan}}\affiliation{\affilf}
\author{Björn \surname{Kruse}}\affiliation{\affilg}
\author{Asbjørn Ø. \surname{Lægdsmand}}\affiliation{\affilk}
\author{Bruno \surname{Langbehn}}\affiliation{\affill}
\author{Suddhasattwa \surname{Mandal}}\affiliation{\affile}
\author{Tommaso \surname{Mazza}}\affiliation{\affild}
\author{Michael \surname{Meyer}}\affiliation{\affild}
\author{Christian \surname{Peltz}}\affiliation{\affilg}
\author{Thomas \surname{Pfeifer}}\affiliation{\affilm}
\author{Safi \surname{Rafie-Zinedine }}\affiliation{\affild}
\author{Antoine \surname{Sarracini}}\affiliation{\affilb}
\author{Mario \surname{Sauppe}}\affiliation{\affila}
\author{Florian \surname{Schenk}}\affiliation{\affiln}
\author{Kirsten \surname{Schnorr}}\affiliation{\affilb}
\author{Björn \surname{Senfftleben}}\affiliation{\affild}
\author{Keshav \surname{Sishodia}}\affiliation{\affilf}
\author{Frank \surname{Stienkemeier}}\affiliation{\affilo}
\author{Zhibin \surname{Sun}}\affiliation{\affilb}
\author{Rico Mayro P. \surname{Tanyag}}\affiliation{\affilp}
\author{Paul \surname{Tümmler}}\affiliation{\affilg}
\author{Sergey \surname{Usenko}}\affiliation{\affild}
\author{Carl Frederic \surname{Ussling}}\affiliation{\affila}
\author{Vanessa \surname{Wood}}\affiliation{\affiln}
\author{Xinhua \surname{Xie}}\affiliation{\affilb}
\author{Maksym \surname{Yarema}}\affiliation{\affiln}
\author{Olesya \surname{Yarema}}\affiliation{\affiln}
\author{Nuri \surname{Yazdani}}\affiliation{\affiln}
\author{Hankai \surname{Zhang}}\affiliation{\affilb}
\author{Bernd \surname{von Issendorff }}\affiliation{\affilo}
\author{Yevheniy \surname{Ovcharenko}}\affiliation{\affild}
\author{Marcel \surname{Mudrich}}\affiliation{\affilk}
\author{Daniela \surname{Rupp}}\email[]{ruppda@phys.ethz.ch}\affiliation{\affila}
\author{Alessandro \surname{Colombo}}\email[]{alcolombo@phys.ethz.ch}\affiliation{\affila}

\begin{abstract}
\newtext{We experimentally demonstrate that pairs of time-delayed ultrabright and ultrashort X-ray pulses of two different colors, delivered by modern X-ray Free Electron Lasers, can provide two time-delayed snapshots of a sample. We introduce \emph{Dichography}, a method that algorithmically separates the diffraction signals overlapping on the detector and independently retrieves the two images of the specimen. We employ \emph{Dichography} to reconstruct two views of individual xenon-doped helium nanodroplets with \SI{20}{\nano\meter} spatial resolution. 
The consistency of structures observed in both images at delays up to \SI{750}{\femto\second} provides evidence that, under these illumination conditions, significant structural damage only occurs at longer timescales.
We further validate the method by imaging pairs of silver nanoparticles intercepted by the same light pulse.} \emph{Dichography} enables a new class of experiments across physics, chemistry, and materials science, making a significant step toward the original promise of X-ray free-electron lasers to capture ultrafast movies of nanomatter.
\end{abstract}
\maketitle


Diffraction experiments are a central method to study ultrafast structural changes in matter when combined with the ultra-short and ultra-bright coherent light pulses delivered by extreme ultraviolet and X-ray Free Electron Lasers (XFELs). XFELs can nowadays deliver individual pulses with a duration of a few tens of femtoseconds down to hundreds of attoseconds \cite{guo2024experimental, franz2024terawatt, yan2024terawatt, maroju2021complex, duris2020tunable}. 
The light diffracted from the sample, recorded by a suitable detector, provides a high-resolution structural fingerprint of the system under study, with a time resolution equivalent to the duration of the light pulse.

Among diffraction-based techniques, single-particle Coherent Diffraction Imaging (CDI) stands out for its capability to retrieve the full electronic density of isolated nanostructures from a single diffraction image produced by an individual light pulse \cite{kirian2020imaging,colombo2023imaging}. Thus, since the pioneering era of XFELs \cite{chapman2006femtosecond}, CDI has been an invaluable tool for capturing ultrafast snapshots of fragile nanoscale objects in free-flight \cite{kimura2014imaging, seibertSingleMimivirusParticles2011, ekeberg2024observation, colombo2023three, langbehnThreeDimensionalShapesSpinning2018} and their light-driven dynamics \cite{gorkhoverFemtosecondAN2016, fluckigerTimeresolvedXrayImaging2016, langbehnDiffractionImagingLight2022, bacellarAnisotropicSurfaceBroadening2022, dold2025melting}.

Dynamical investigations are mostly performed with \emph{pump-probe} schemes: processes in the system are triggered by a first laser pulse -- the \emph{pump} -- and its structural and electronic properties are probed by an XFEL flash that interacted with the sample at a specific time delay -- the \emph{probe}. Often, the \emph{pump} laser covers spectral regions from the near infrared to the ultraviolet. 
At the corresponding photon energies of a few \unit{\electronvolt}, only a subset of the possible light-induced ultrafast processes can be triggered and, thus, investigated.
Furthermore, the \emph{pump} doesn't provide any structural information on the sample, whose features are at the nanometer length-scale -- well beyond the resolution limit of the UV or longer-wavelength laser radiation.

Thanks to recent technical developments, XFELs can deliver pairs of light pulses in the XUV or X-ray spectrum \cite{serkez2020opportunities,guo2024experimental,lutman2013experimental, marinelli2015, allaria2013two, hara2013two, prat2022widely}. Both pulses retain the coherence, brightness, and ultrashort nature of conventional XFEL pulses; their wavelength can be independently tuned, and their arrival time at the interaction region can be controlled with high precision. This novel capability, known as \emph{two-color} mode, unlocks a whole new class of dynamical investigations, where the XFEL delivers both the \emph{pump} pulse, which triggers the dynamics in a pristine sample, and the \emph{probe} pulse, which probes the evolved state. 
When the \emph{two-color} capability of XFELs is combined with the high spatial resolution of diffraction imaging, the sample's structure can be accessed at two different points in time.
The femtosecond precision at which the time delay between the two light pulses can be controlled paves the way for two-frame movies of structural dynamics approaching the Petahertz realm, and potentially well beyond. 

However, light detectors are several orders of magnitude slower \cite{struder2010, kuster2021,hinger2022advancing}. They cannot be triggered or read-out quickly enough to individually capture the diffraction of the \emph{pump} and the \emph{probe} shots. As a consequence, the sum of their signal is recorded in a single diffraction image.
Conventional algorithmic methods for CDI, necessary to restore the sample's image from the acquired scattering data \cite{kirian2020imaging, colombo2023imaging, marchesini2003x} are not applicable.

Significant effort has been devoted to overcoming this technical limitation by physically separating the two scattering images. First successful realizations achieved this by placing spectral filters~\cite{ferguson2016,hecht2018b} or by directing the pulses onto two different detectors via optical devices~\cite{sauppe2024double} \newtext{or different regions of the detector through holographic approaches~\cite{gunther2011sequential}}. Nevertheless, these solutions come with significant trade-offs, e.g., strongly reduced signal brightness, restricted range of viable photon energies, reduced control on the time delay, \newtext{partial loss of scattering information or limited spatial resolution}.
Therefore, developing new methods to fully exploit \emph{two-color} pulses across various diffraction-based experimental approaches remains a top priority for the scientific community.

In this work, we introduce \emph{Dichography} (from \emph{dichos}, meaning ``in two''), an imaging method that enables the reconstruction of two sample images from superimposed diffraction signals \newtext{without any prior knowledge of the samples shape and size}.
\newtext{While relying on an established numerical and theoretical framework \cite{millane2015phase}, \emph{Dichography} extends the approach to real experimental conditions, i.e., samples with unknown structures, noisy diffraction images and missing information in the center of the patterns. Thereby, it unlocks the first experimental demonstration of single-particle two-color Coherent Diffraction Imaging at XFELs, disentangling the scattering contributions in the recorded data and reconstructing two distinct \emph{views} of the sample as “seen” by the individual X-ray colors.}

\newtext{Superfluid helium nanodroplets doped with xenon are intercepted by two-color X-ray pulses delivered by the European XFEL at photon energies of \pump and \probe.} Despite the low photon yield in the diffraction signal for such a highly photon-demanding approach, successful reconstructions can be achieved for a subset of optimal diffraction patterns. The retrieved \emph{frames} are two snapshots of the same xenon-doped helium droplet, delayed in time by \newtext{\SI{50}{\femto\second} and \SI{750}{\femto\second}.} 
\newtext{The retrieved embedded xenon structures appear to be unchanged within the spatial resolution of the experiment, indicating that extensive structural damage of the helium-embedded nanostructures due to the interaction with the first XFEL pulse only occurs at longer time scales.}

\newtext{For benchmarking the \emph{Dichography} method, a further experimental case is also presented. Here, }pairs of silver nanoparticles are simultaneously present in the focus when single light pulses delivered by the SwissFEL reach the interaction region. When the two samples are sufficiently separated in space, the interference of their scattered fields is not resolved by the detector. The properties of the imaging problem, i.e., the mathematical connection between the sample and the measured diffraction, are analogous to the \emph{two-color} situation\newtext{, i.e., an incoherent superposition of two diffraction patterns}. Therefore, \emph{Dichography} can be employed to restore the images of the two particles.
\newtext{While the analysis of these \emph{double-hit} patterns is of limited scope regarding the physical implications, it serves as a crucial testbed, allowing us to show and investigate the performance of the method on experimental data for optimal diffraction conditions, not affected by the limited pulse energies of XFELs in \emph{two-color} operation mode, and for diverse structural properties of the samples.}

The imaging quality and resolution achieved by \emph{Dichography} is strictly connected to the present understanding of the underlying mathematical and numerical challenges, as well as to the current performance of XFELs. With this work, we also aim at stimulating and encouraging research efforts in both directions.
In this regard, the possible applications of \emph{Dichography} with the currently available experimental possibilities are discussed, along with concepts that could be viable in the near future. 

The demonstration of \emph{Dichography}, and its possible improvements and extensions to closely related imaging challenges, unlock novel and exciting possibilities for the study of ultrafast structural dynamics resolved in both time and space.

\section{The Dichographic problem}\label{sec:dichoprob}

\emph{Dichography}, similar to conventional CDI, is an \emph{indirect} imaging method. The sample image is retrieved from the acquired diffraction data via algorithmic methods, designed to solve the specific \emph{imaging problem}.

In conventional CDI, the image recorded by the detector, $I(\vec{q})$, is the square amplitude of the scattered field $\psi(\vec{q})$. The field carries the full information on the sample density $\rho(\vec{x})$. However, only its magnitude $\left| \psi(\vec{q})\right| = \sqrt{I(\vec{q})}$ is experimentally accessible, while the field's phase is lost. Therefore, it is necessary to retrieve the phase to obtain the sample image $\rho(\vec{x})$. \newtext{In two dimensions, the image $\rho(\vec{x})$ is proportional to the projection of the electronic density along the beam propagation direction \cite{kirian2020imaging}}.
In CDI, the \emph{phase retrieval problem} is numerically solved by specifically designed \emph{phase retrieval algorithms} \cite{marchesini2007invited,colombo2023imaging}.

The imaging problem in \emph{Dichography} is significantly more demanding. Indeed, \emph{Dichography} deals with situations where the acquired scattering data $I(q)$ is the incoherent sum of the intensities of two independent scattered fields $\psi^\one(\vec{q})$ and $\psi^\two(\vec{q})$, that is:
\begin{equation}\label{eq:dichocore}
I(\vec{q}) = I^\one(\vec{q}) + I^\two(\vec{q}) = \left| \psi^\one(\vec{q})\right|^2  + \left| \psi^\two(\vec{q})\right|^2 \,.
\end{equation}
The two fields encode two different electronic densities $\rho^\one$ and $\rho^\two$, which correspond to the two independent images of a sample. Thus, the \emph{dichographic} imaging problem consists of retrieving the phase of $\psi^\one(\vec{q})$ and $\psi^\two(\vec{q})$, as well as their relative amplitude contribution to the recorded signal $I(\vec{q})$.

\begin{figure}
\includegraphics[width=\columnwidth]{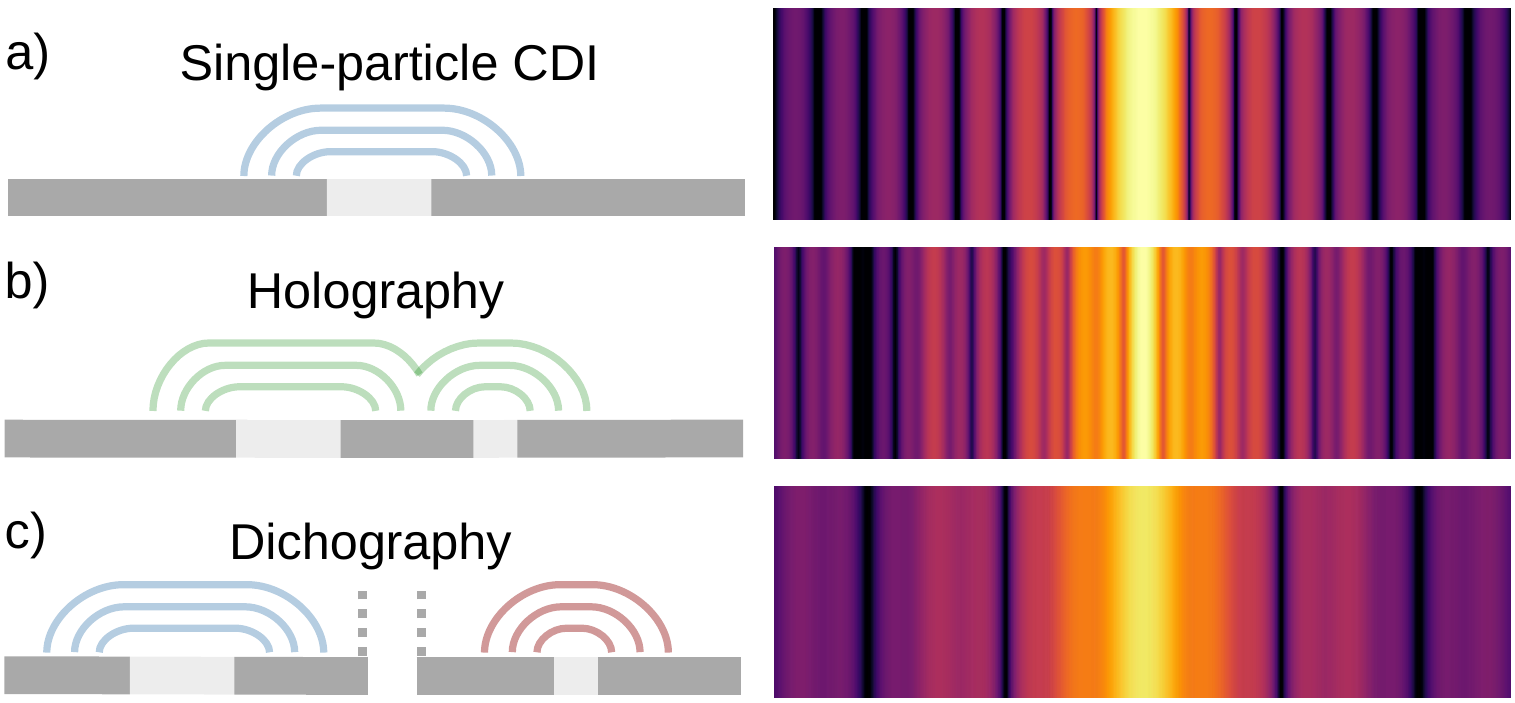}
\caption{Intuitive representation of the difference between conventional single-particle CDI, \emph{Holography}, and \emph{Dichography} in terms of a double-slit experiment. In a), the incoming radiation intercepts a single slit, and the recorded scattering image encodes its size. In b), the fields scattered by two apertures interfere with each other. The \emph{far-field} intensities thus encode the information on the size of both slits and their relative distance. In c), the fields produced by the two apertures do not interfere. The diffraction signal, thus, only encodes the independent properties of the two slits.}
\label{fig:holocomp}
\end{figure}

Fig.~\ref{fig:holocomp} provides an intuitive comparison between \emph{Dichography}, single-particle CDI and \emph{Holography}, using a double-slit diffraction experiment as a reference.
As previously mentioned, conventional CDI, as illustrated in Fig.~\ref{fig:holocomp}a, addresses the reconstruction of a single isolated sample, denoted by $\rho$. 
In contrast, \emph{Holography}, depicted in Fig.~\ref{fig:holocomp}b, involves two distinct samples, $\rho^\one$ and $\rho^\two$. This configuration necessitates the interference of the two resulting scattering fields at the detector, which fundamentally differs from the \emph{Dichography} approach shown in Fig.~\ref{fig:holocomp}c.

While \emph{Dichography} is designed to deal with \emph{two-color} diffraction images, its applicability further extends to diverse experimental schemes where the link between the recorded data and the samples under study follows Eq. \eqref{eq:dichocore}. For this reason, we can generically address $\rho^\one$ and $\rho^\two$ as the two \emph{frames} of a \emph{dichographic} reconstruction. The two \emph{frames} are, conceptually, independent from each other and can be structurally unrelated.

The imaging algorithm at the core of \emph{Dichography}, which disentangles the scattering contributions and retrieves the two frames, is a fundamental result of this work. 
\newtext{The possibility of algorithmically disentangling scattering contributions incoherently adding up on the detector has already been shown in previous works \cite{millane2015phase}, stimulated by the problem of imaging for arrangements beyond crystallography \cite{elser2008reconstruction}. \emph{Dichography} extends this concept to the reconstructions of isolated samples with unknown structures, thus effectively enabling two-color single-particle coherent diffraction imaging.
The \emph{dichographic} imaging problem is less constrained than conventional CDI and, thus, inherently \emph{harder}. Still, in most practical cases, it is demonstrated to have a unique solution \cite{millane2015phase}, as further discussed in Sec.~\ref{sec:further}.}

\newtext{The algorithmic} structure is derived from that of conventional \emph{iterative phase retrieval algorithms} for single-particle CDI \cite{fienup1982phase, marchesini2007invited, elser2003phase}, adapted to accommodate the specific characteristics of \emph{Dichography}\newtext{, extending the scheme proposed in Ref.~\cite{millane2015phase} to the retrieval of samples with unknown shapes}. 

A detailed description of the algorithm's workflow and its connection to the CDI counterpart is deferred to \methods~\ref{sec:dichoalg}. We now turn to an investigation of the performance of \emph{Dichography} in two different experimental situations.

\section{Results}\label{sec:results}

%
%

\subsection{\newtext{Two-color Coherent Diffraction Movies of Xenon-doped Helium Nanodroplets}}\label{sec:xedata}

\begin{figure*}
\includegraphics[width=\textwidth]{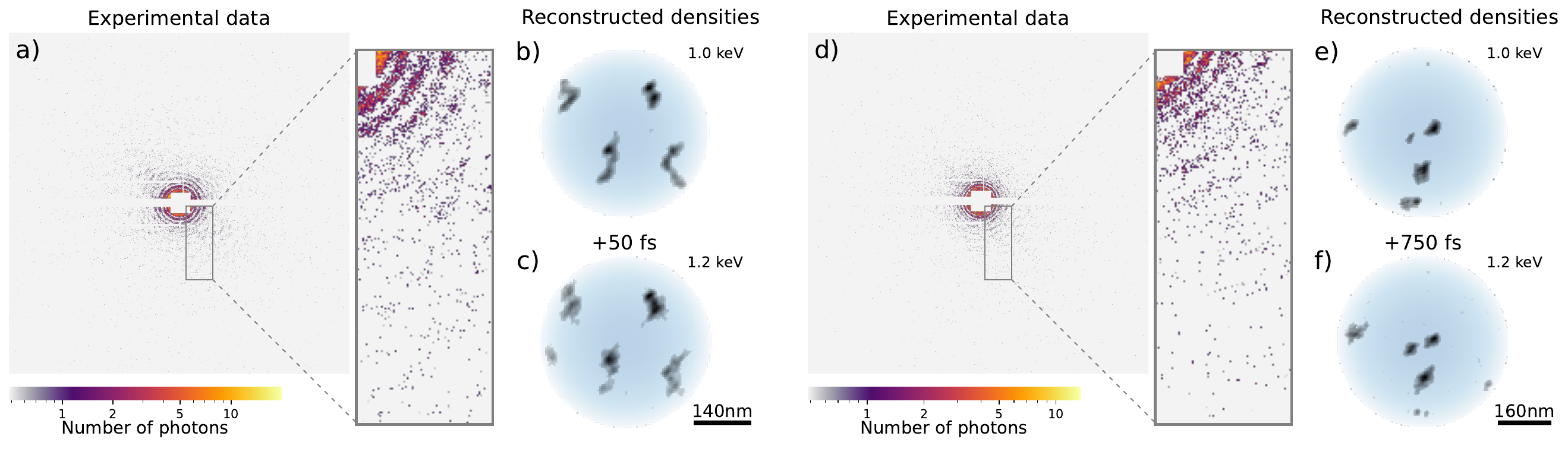}
\caption{\newtext{Reconstructions from two-color diffraction patterns acquired at the European XFEL, produced by superfluid helium nanodroplets doped with xenon. Two experimental diffraction patterns are reported in a) and d), on which the diffraction signals produced by two different photon energies, \pump and \probe, are superimposed. 
The intensity is encoded as reported by the color bar at the bottom. The intensity values are translated into numbers of \SI{1.2}{\kilo\electronvolt} photons per pixel. The inset plot is a zoomed-in region of the diffraction image, to better appreciate the photon statistics, and thus the statistical noise, that affects the experimental data. 
The two colors are delayed by \SI{50}{\femto\second} for the pattern in a) and \SI{750}{\femto\second} for the pattern in d).
The two reconstructed frames, corresponding to the two time-delayed projections of the particle density, are reported in b), c) and e), f) for the patterns in a) and d), respectively. 
The reconstructed density of the xenon doping follows the color bar in Fig.~\ref{fig:SFEL_examples}. The shape of the helium droplets, independently retrieved before the imaging process and constrained using the DCDI method, are superimposed in blue color.}}
\label{fig:SQS_results}
\end{figure*}

The analysis of diffraction patterns produced by two-color XFEL light pulses is the fundamental motivation behind the development of \emph{Dichography}. The two light pulses, interacting at different times with the sample, imprint its time evolution in the diffraction signal. The two \emph{views} on the sample are reconstructed by \emph{Dichography} to gain a full-fledged two-frame movie. However, this intriguing potential comes along with technical challenges, as discussed in the following.

The experiment was performed at the European XFEL facility, taking advantage of the recently available capability of producing two collinear X-ray pulses of different radiation wavelengths with controllable time delay \cite{serkez2020opportunities}. The two co-propagating pulses are tuned to \SI{1.0}{\kilo\electronvolt} and \SI{1.2}{\kilo\electronvolt} photon energy, corresponding to \SI{1.2}{\nano\meter} and \SI{1.0}{\nano\meter} wavelength, and can be separated in time \newtext{from  \SI{50}{\femto\second} up to \SI{750}{\femto\second}}. Both are focused in the interaction region, where they intercept isolated superfluid helium nanodroplets doped with xenon. An individual droplet in the focus interacts with both pulses, and the corresponding scattered fields incoherently overlap on the detector. Further details on the experiment are discussed in \methods \ref{subsec:SQS}.

Two diffraction images acquired in these conditions are shown in Fig.~\ref{fig:SQS_results}a and Fig.~\ref{fig:SQS_results}d \newtext{for two different time delays between the two colors, i.e., \SI{50}{\femto\second} and \SI{750}{\femto\second}, respectively}. 
A relevant feature of these patterns is the low number of photons recorded by the detector, if compared to the photon yield for the standard, single pulse, operation mode of the XFEL \cite{hecht2025two} (see \methods~\ref{subsec:SQS}). 
This aspect represents a challenge for a photon-hungry technique such as CDI, and thus even more so for \emph{Dichography} (as further investigated in Sec.~S1 of the Supplemental Material~ \cite{supplemental}).

Nevertheless, the specific properties of the physical system under study, i.e., doped superfluid helium nanodroplets, allow us to take advantage of the Droplet Coherent Diffraction Imaging (DCDI) approach \cite{tanyag2015communication}. The DCDI method constrains the underlying helium droplet structure during the actual reconstruction process. The droplet size and brightness are extracted in the pre-processing phase, by fitting the experimental data with the diffraction profile of a spherical sample.
The information to be retrieved, therefore, pertains solely to the internal structure of the doping material. For conventional CDI, this greatly reduces the complexity of the problem, it improves the convergence properties of the algorithms compared to standard CDI, and it makes the reconstruction process much more resilient to noise \cite{tanyag2015communication}. 

The DCDI method has been incorporated into the \emph{Dichography} reconstruction process through a straightforward adaptation of its original workflow, as discussed in \methods~\ref{sec:fitting}, and is used here to analyze the two-color diffraction images presented in this section. Its applicability to this dataset relies on the possibility of extracting the droplet size as well as on the persistence of the pristine helium droplet density, embedding the xenon doping, up to a time delay of \SI{750}{\femto\second}. Both aspects are investigated and verified using pure-helium data acquired during the same experimental campaign, as reported in Ref.~\cite{hecht2025two}.

Despite the demanding experimental conditions, the reduction in problem complexity achieved through the use of DCDI makes these two-color data tractable with \emph{Dichography} in a few optimal cases.

\newtext{
The two frames reconstructed from the diffraction pattern shown in Fig.~\ref{fig:SQS_results}a are presented in Fig.~\ref{fig:SQS_results}b for the \pump pulse, and in Fig.~\ref{fig:SQS_results}c for the following \probe pulse. The two reconstructions are thus two snapshots of the same nanoparticle delayed by \SI{50}{\femto\second}.
In a similar way, the two frames reconstructed from the pattern in Fig.~\ref{fig:SQS_results}d correspond to the snapshots of the initial particle density (Fig.~\ref{fig:SQS_results}e) and its density after \SI{750}{\femto\second} (Fig.~\ref{fig:SQS_results}f).}

\newtext{
In the reconstructions reported in Fig. ~\ref{fig:SQS_results}, high density regions arise, structurally compatible with previous observations of xenon structures inside superfluid helium nanodroplets \cite{gessner2019imaging, tanyag2015communication,ulmer2023generation}. 
The first frame in Fig.~\ref{fig:SQS_results}b reveals the presence of four xenon agglomerates. The same structures can be identified in the second frame in Fig.~\ref{fig:SQS_results}b. However, the \probe reconstruction is characterized by a visibly lower quality, with ``fuzzier'' xenon agglomerates and the presence of a high density region on the left side of the droplet, which is not present in the \pump reconstruction. The lower quality of the \probe reconstruction can be explained with the imbalance of the scattering contributions: indeed, the recorded intensity of the scattered \probe signal is less than half of the \pump contribution.}

\newtext{
A similar quality of the two reconstructed frames is observed in the second example. Here, both the \pump and the \probe reconstructions (Fig.~\ref{fig:SQS_results}e and Fig.~\ref{fig:SQS_results}f, respectively) show the same high density spots extending from the bottom of the particle towards the center, in addition to a dense spot on the left edge. Few low density features, not compatible between the two frames, can be spotted. Differently from the \SI{50}{\femto\second} reconstruction, here the two frames show similar quality in terms of noise and artifacts. In this case, the overall intensity of the diffraction in Fig.~\ref{fig:SQS_results}d, i.e., the total number of recorded photons, is \SI{30}{\percent} lower than the one of the pattern in Fig.~\ref{fig:SQS_results}a. However, the two intensity contributions are significantly more balanced, with a \probe contribution to the scattering just \SI{13}{\percent} stronger than the \pump contribution.}

\newtext{
From the representation of the retrieved densities in Fig.~\ref{fig:SQS_results}, the reader may question whether the similar structures in the two frames arise merely as artifacts, i.e., as a reproduction of the same signal in both frames. Here, we address this possibility as \emph{ghosting}. We further discuss this aspect in the following Sec.~\ref{subsec:SwissFEL} and we provide a mathematical interpretation in Sec.~\ref{subsec:uniqueness}. In the case of two-color diffraction, i.e., of two different radiation wavelengths, the sample is imaged with intrinsically different spatial resolutions in the two frames.} Specifically, the spatial extent corresponding to a single pixel varies between the frames, becoming smaller at higher photon energies due to the increased resolution at a shorter wavelength. As a result, the same structure spans a different number of pixels in the two reconstructions, and appears visually similar in size only when both are displayed with a common scale bar, as done in Fig.~\ref{fig:SQS_results}. This point is further illustrated in Fig.~\ref{fig:SQS_results_scaled} and discussed in more detail in \methods~\ref{sec:fitting}. \emph{Ghosting} in two-color \emph{Dichography} would therefore give rise to a replicated structure but with a different scaling. Thus, its presence can be excluded for the xenon structures observed in Fig.~\ref{fig:SQS_results}.

The presence of the same structure in both frames, apart from the artifacts discussed above, implies that the xenon agglomerates survive for at least \SI{750}{\femto\second} after irradiation with the \emph{pump} pulse. \newtext{Thus, these results suggest that, at the irradiation level reached in this experiment, significant structural damage of the xenon filaments only happens at longer timescales. This observation, which holds up to the achieved resolution, is supported by molecular dynamics simulations \cite{hecht2025two}.}

\newtext{While the geometry of the experiment would allow for a spatial resolution down to \SI{5}{\nano\meter}, the effective resolution achieved by the reconstructions is only around \SI{20}{\nano\meter} (see Fig. S11 of the Supplemental Material). This is mostly due to the low intensity of the patterns that limits the maximum momentum transfer at which useful signal is recorded.}

\newtext{Among all the patterns recorded during the experiment, only a few turned out to have the right conditions for a \emph{dichographic} reconstruction, due to the challenging conditions of the experiment. First of all, the pulse energy of the two-color pulses is reduced due to the specific XFEL configuration (see Sec.~\ref{subsec:SQS}). As a consequence, the brightness of the diffraction patterns is limited. In this low intensity condition, \emph{Dichography} can successfully reconstruct the two frames only if their intensity contributions to the scattering signal are similar, as shown in Sec.~S1.3 of the Supplemental Material~\cite{supplemental}. A similar diffraction contribution requires a similar intensity of the two pulses when they interact with the droplet. This condition is however only achieved in a limited region of the focal volume due to the geometry of the experiment and the focusing optics, as also discussed in Sec.~\ref{subsec:SQS}. As a result, for this experiment, two-color diffraction patterns suitable for \emph{Dichography} are ``rare events'' in the dataset and can be identified with a specific procedure described in Sec. S2 of the Supplemental Material~\cite{supplemental}.}

%

\subsection{Seeing double from a single shot}\label{subsec:SwissFEL}

\begin{figure*}
\includegraphics[width=0.9\textwidth]{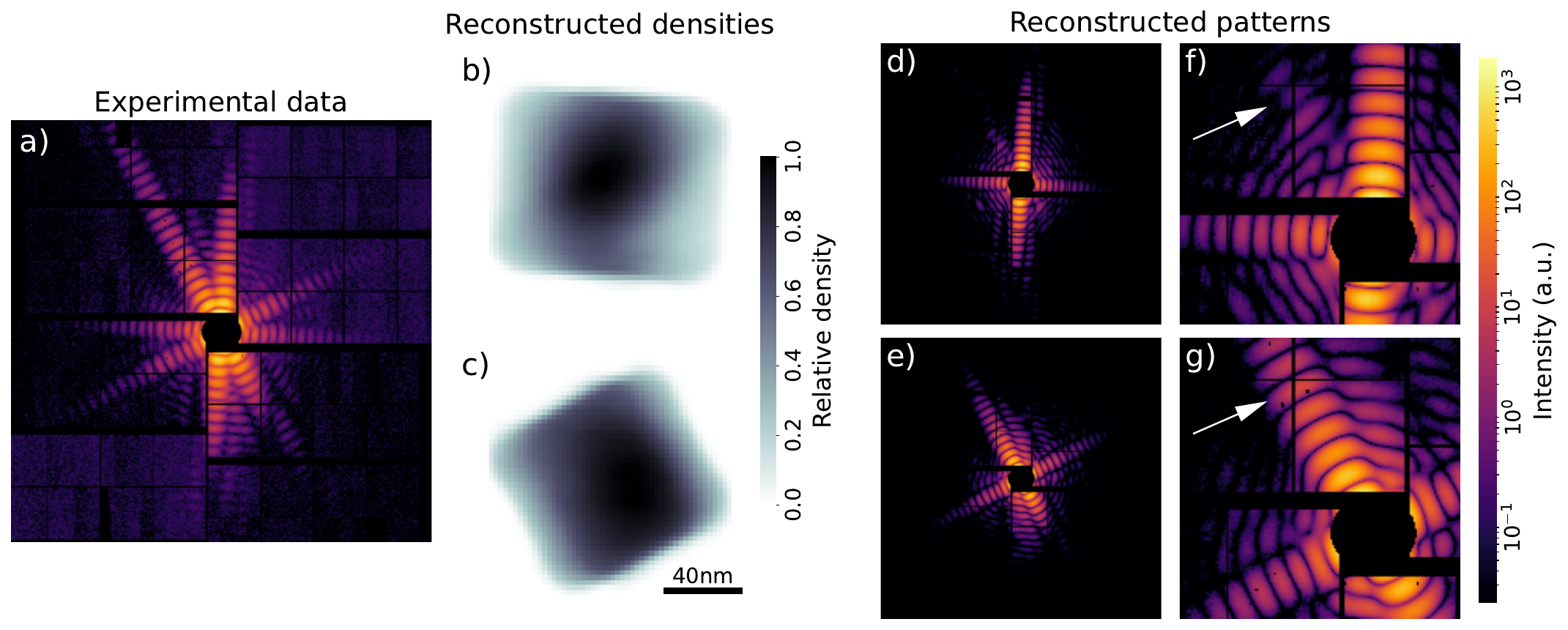}
\caption{Example of \emph{dichographic} reconstruction from a single diffraction pattern produced by a ``double-hit''. The experimental diffraction data from an individual light pulse delivered by SwissFEL is shown in a). The two particle densities retrieved from a) via \emph{Dichography} are shown in b) and c). The two reconstructions correspond to two silver nanocubes, with edge length of around \SI{100}{\nano\meter}. Once the two particles have been reconstructed, the corresponding single-particle diffraction patterns can be calculated, shown in d) for the reconstruction in b), and in e) for the one reported in c). The central parts of the disentangled diffraction images in d) and e) are magnified in f) and g), respectively. The white arrows point at the same region of the diffraction image. All diffraction data in a), d), e), f) and g) are plotted with logarithmic color scale.}
\label{fig:SFEL_single}
\end{figure*}

\newtext{The second set of results concerns the analysis of diffraction patterns produced by two particles struck by the same XFEL shot, known as ``double-hits'', and they serve as a valuable probe for investigating the capabilities, performance, and key features of \emph{Dichography}.}

Here, pairs of silver nanoparticles, simultaneously present in the XFEL focus, are intercepted by a single light pulse delivered by SwissFEL \cite{nolting2023swiss}. When the particles are far enough apart, the interference term between the two scattered fields is not resolved by the detector. The recorded diffraction image is, in practice, the incoherent sum of the two intensities, suitable for imaging with \emph{Dichography}. The two retrieved \emph{frames} correspond to the electron density of the two different particles. Further experimental information is provided in \methods \ref{subsec:SwissFELExp}.

\newtext{A first \emph{double-hit} reconstruction} is shown in Fig.~\ref{fig:SFEL_single}. The images of the two particles in Fig.~\ref{fig:SFEL_single}b and \ref{fig:SFEL_single}c are the two frames retrieved from the experimental diffraction pattern reported in Fig.~\ref{fig:SFEL_single}a, without prior information regarding the particles’ shape, size, or orientation. 
The reconstructions reveal that the experimental image is produced by two silver nanocubes of slightly different sizes and alignment. Due to the cubic structure of the two samples and their orientation, each cube produces strong streaks with four-fold symmetry in the diffraction image. A total of eight streaks can be indeed identified in the diffraction pattern in Fig.~\ref{fig:SFEL_single}a. 

It is possible to disentangle the two contributions to the scattering, i.e., to calculate the corresponding single-particle diffraction images from the two reconstructed frames. Fig.~\ref{fig:SFEL_single}d and Fig.~\ref{fig:SFEL_single}e report the two disentangled scattering contributions from the first (Fig.~\ref{fig:SFEL_single}b) and second (Fig.~\ref{fig:SFEL_single}c) frames, respectively. There, we can appreciate the capability of \emph{Dichography} to separate the two contributions to the scattering superimposed in \ref{fig:SFEL_single}a, by reproducing the four-fold symmetry of the two cubes with different orientations. 
This situation is further muddled by a closer view in the central part of the separated patterns where the strongest contribution to the scattering signal is concentrated, provided in Fig.~\ref{fig:SFEL_single}f and \ref{fig:SFEL_single}g.

From these zoomed-in representations of the disentangled scattering signals, it is also possible to identify few regions where the intensities are not fully disentangled. For example, the white arrows in Fig.~\ref{fig:SFEL_single}f and Fig.~\ref{fig:SFEL_single}g indicate the same coordinates in the two patterns.
In Fig.~\ref{fig:SFEL_single}g, this area corresponds to a high intensity streak in the direction of cube facets, as visible from the reconstruction in Fig.~\ref{fig:SFEL_single}c.
Despite the absence of any related structural feature in the other reconstructed frame in Fig.~\ref{fig:SFEL_single}b, its disentangled pattern in Fig.~\ref{fig:SFEL_single}f still presents traces of this strong scattering signal. This signal marginally affects the inner density profile in Fig.~\ref{fig:SFEL_single}c, slightly distorted in the direction of the cube facets of the other frame. We refer to this ``cross-talking'' between the two frames -- where features in one create slight artifacts in the other due to strongly differing scattering intensity in specific directions -- as \emph{ghosting}. \newtext{The origin of ghosting artifacts is further discussed in Sec.~\ref{subsec:uniqueness}.} 
On the other hand, the intensities mistakenly assigned to the first frame in Fig.~\ref{fig:SFEL_single}f are about two orders of magnitude lower than those correctly retrieved in Fig.~\ref{fig:SFEL_single}g. This is evident from the logarithmic color map on the right and underscores the ability of \emph{Dichography} to separate the two signals, even when their intensities \newtext{locally differ by orders of magnitude in the diffraction signal}.

\begin{figure}
\includegraphics[width=\columnwidth]{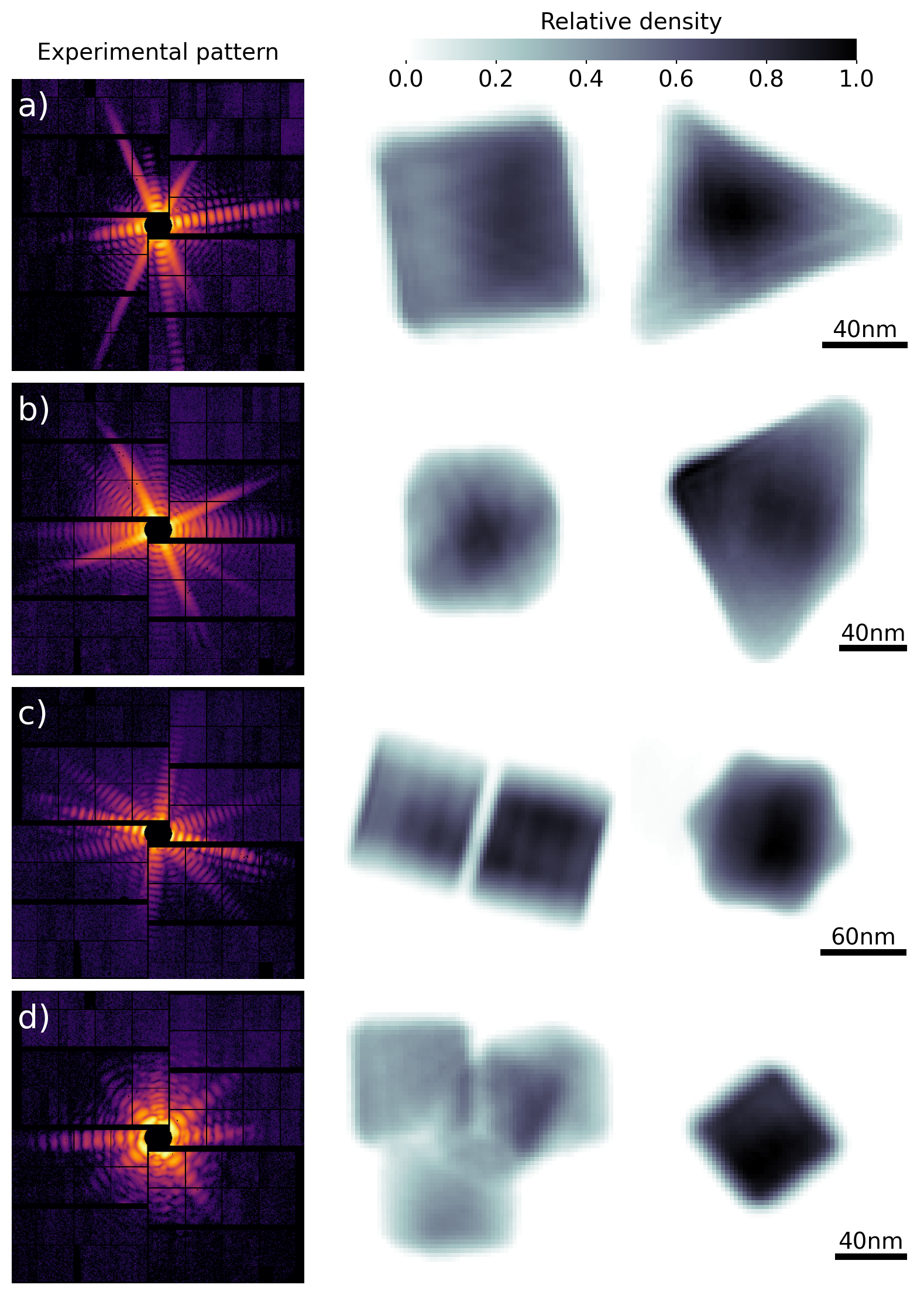}
\caption{Further \emph{Dichography} reconstructions of silver nanoparticles from individual diffraction patterns acquired at SwissFEL. Each sub-figure, from a) to d), reports different imaging results on a single diffraction pattern. The experimental data are shown in the leftmost column. The central and rightmost columns report the two frames of the reconstruction. Similar to Fig.~\ref{fig:SFEL_single}, within the same sub-figure the two reconstructed densities are reported with common color scale and scale bar, such that their size and brightness can be compared.}
\label{fig:SFEL_examples}
\end{figure}

Further reconstruction attempts on diffraction patterns extracted from the same dataset of Fig.~\ref{fig:SFEL_single} are presented in Fig.~\ref{fig:SFEL_examples}. These examples are selected to highlight how \emph{Dichography} performs on different sample configurations. 
For instance, Fig.~\ref{fig:SFEL_examples}a shows a reconstruction which reveals a pair of nanostructures with completely different geometries, a cube and a triangle, strongly underlining how this method neither requires any specific constraint in terms of sample's structure, nor structural links between the two frames.
The reconstruction in Fig.~\ref{fig:SFEL_examples}b has an additional challenging aspect. Here, the spatial extension of the two samples differs significantly, with the smaller one being around \SI{100}{\nano\meter} in maximum size and the second reaching around \SI{200}{\nano\meter}. As a consequence, the smaller sample could spatially ``fit'' inside the larger one, potentially rendering the separation of the two frames of the reconstruction a significantly harder task. 
In both Fig.~\ref{fig:SFEL_examples}a and Fig.~\ref{fig:SFEL_examples}b, \emph{ghosting} artifacts are barely visible in the form of line-shaped density fluctuations aligned with the hard edges of the reconstruction in the other frame. 

Another particular case is shown in Fig.~\ref{fig:SFEL_examples}c. Here, one sample has a highly symmetric hexagonal morphology, while the other reconstructed density is an agglomerate of two cubes. 
The apparent empty space between the two cubes, visible due to the accidental perfect alignment of their faces with the FEL pulse, is actually caused by the residual material that coats the nanoparticles in the buffer solution. This buffer is crucial for maintaining stable aerosol injections and ensuring the colloidal stability of the nanoparticles.
Thanks to a scattering cross-section significantly lower than silver, it appears mostly transparent in CDI reconstructions. Despite being two separated particles, the two nanocubes are reconstructed together in the same frame with their relative placement, as their interference signal is resolved by the detector. The spatial separation of the hexagonal particle from the cube pairs is, instead,  large enough to render the interference term undetected. Therefore, it is reconstructed by \emph{Dichography} in a separate frame.

The last example on ``double-hit'' data is reported in Fig.~\ref{fig:SFEL_examples}d, and presents many peculiarities of the samples shown before combined in one case. The two frames of the reconstruction have a significantly different spatial extension, \SI{80}{\nano\meter} for the smaller and up to a maximum of \SI{200}{\nano\meter} for the larger. The smaller density has a simple cubic architecture, whereas the larger one can be interpreted as an agglomerate of four nanocubes. The shape of three cubes in the agglomerate is well imprinted into the overall outline, while the presence of the fourth cube is only revealed thanks to the successful reconstruction of the density values inside the sample. 
Furthermore, the inner fourth cube in the agglomerate does not originate from a \emph{ghosting} artifact of the cube in the other frame: their orientation is, indeed, slightly different.

The results in Fig.~\ref{fig:SFEL_single} and Fig.~\ref{fig:SFEL_examples} discussed so far demonstrate that \emph{Dichography} can deal with markedly different shapes and sizes between the two retrieved frames. 
An additional relevant feature of single-particle CDI experiments at XFELs are the fluctuations in the brightness of the diffraction signals. In fact, samples intercept the light pulse while free-flying, and their position in the focus at the diffraction event is ruled by statistics. Particles can intercept the light pulse at varied positions in the focal volume, with distinct light intensity and, consequently, diffraction brightness. This is further enhanced by intensity fluctuations in the XFEL source.

It is straightforward to deduce that, in the ``double-hit'' case, two particles far apart in the interaction region can be subjected to different pulse intensities, causing both deviations in the overall brightness as well as imbalance in the scattering contribution between the two frames.
Less intuitively, this is also a key feature in the \emph{two-color} case, due to the different positions of the foci of the two pulses, as discussed in the next section (see Sec. \ref{subsec:SQS} for technical details).
These factors must be considered when evaluating the imaging performance of \emph{Dichography}.

Imaging results on double silver nanoparticles provide a valuable opportunity to comment on the method's performance in this regard. 
In particular, brightness balancing can be estimated for successful reconstructions, for which we can separate the individual contributions to the scattering signal (as shown in Fig.~\ref{fig:SFEL_single}d and \ref{fig:SFEL_single}e). For the example reported in Fig.~\ref{fig:SFEL_single}, the ratio between the total number of recorded photons is very close to one, with the first frame (Fig.~\ref{fig:SFEL_single}b) yielding just \SI{35}{\percent} more photons than the second one (Fig.~\ref{fig:SFEL_single}c). The asymmetry in scattered photons increases significantly for other cases. The number of photons scattered by the first frame in \newtext{Fig.~\ref{fig:SFEL_examples}c} turns out to be 3.3 times higher than the second one, while the ratio goes up to a factor 3.8 times for the reconstruction in \newtext{Fig.~\ref{fig:SFEL_examples}b}.

The results on ``double-hits'' discussed so far represent a strong demonstration of the viability of \emph{Dichography} for significantly differing morphologies and sample properties. This dataset is characterized by optimal brightness conditions, with the XFEL operating around its maximum performance in terms of total power per pulse and intensity in the focus. 


\section{Discussion}\label{sec:discussion}

In this work, we introduce the \emph{Dichography} method to perform Coherent Diffraction Imaging (CDI) in cases were two non-interfering diffraction signals overlap on the detector. This method finds direct application in \emph{two-color} experiments at XFELs, where pairs of ultrashort pulses with controllable time delay and different wavelengths can be generated. \emph{Dichography} enables the reconstruction of two independent frames, corresponding to the two distinct \emph{views} of the system provided by the different colors, effectively unlocking the possibility of producing a full-fledged movie of the sample\newtext{, without a-priori knowledge}.

The applicability of \emph{Dichography} extends beyond the two-color case, as it generally applies to any scenario where the recorded diffraction data is the incoherent sum of two independent signals \newtext{ \cite{millane2015phase}}. The so-called ``double-hit'' case, where two particles, spatially separated in the interaction region, are simultaneously illuminated by a single light pulse, is one such scenario and serves as a benchmark for the method.

The various examples provided (see Fig.~\ref{fig:SFEL_single} and Fig.~\ref{fig:SFEL_examples}) cover a wide range of cases in terms of sample shape, spatial extent and brightness of the two scattering signals. They serve as practical demonstrations that, in principle, no prior knowledge about the samples is required, nor are any specific constraints imposed on the relationship between the properties of the two retrieved frames.
\newtext{The only assumption about the samples is their isolation which, similar to conventional single-particle CDI, ensures the \emph{oversampling} condition.}

The brightness of the patterns and the distribution of the photons between the two frames become significantly more restrictive for the \emph{two-color} imaging case, were the two overlapping fields come from two XFEL pulses at different photon energies. Here, the special mode of operation of the XFEL undulators delivers two light pulses with significantly reduced intensity with respect to standard operation mode~\cite{serkez2020opportunities}, as further discussed in \methods \ref{subsec:SQS}. Therefore, the reconstructions of xenon-doped superfluid helium nanodroplets shown in Fig.~\ref{fig:SQS_results} are only possible by employing the DCDI method \cite{tanyag2015communication}, which is restricted to samples embedded in nanoparticles with known shape and low scattering cross-section -- so far only demonstrated for the specific case of superfluid helium nanodroplets \cite{tanyagCHAPTERExperimentsLarge2017}. 

The employment of the DCDI approach greatly reduces the problem complexity and makes the reconstruction process significantly more resilient to noise. \newtext{Under the experimental conditions of our measurement,} only specific patterns with optimal condition, i.e., higher brightness and similar photon yields between the two frames, turned out to be tractable \newtext{(as further discussed in Sec. S2 of the Supplemental Material \cite{supplemental})}.

Tests on simulated two-color diffraction images\newtext{, however, suggest that even just a twofold increase of the diffraction intensity can lead to a radical improvement in the reconstruction quality and reliability (see Sec. S1.4 of the Supplemental Material \cite{supplemental}).}

\emph{Dichography}, similar to conventional single-particle CDI, is an \emph{indirect} imaging technique, as the sample image has to be retrieved algorithmically from the measured data \cite{kirian2020imaging, colombo2023imaging}. The algorithmic approach at the core of \emph{Dichography} is built on a direct, relatively straightforward, adaptation of the algorithms for single-shot CDI, as discussed in \methods \ref{sec:dichoalg}. The higher complexity of the \emph{dichographic} problem and its peculiarities render the algorithm less effective in the reconstruction process. Therefore, the imaging performance is enhanced by coupling with an \emph{evolutionary algorithm}, which already demonstrated superior performance and reliability for conventional single-particle CDI \cite{colombo2025spring}, as further discussed in \methods~\ref{sec:MPR}.
A more rigorous and fundamental understanding of the problem, which substantially differs from conventional CDI in some key aspects, could result in the design of more efficient imaging algorithms and analysis workflows, extending the range of applicability even further.

\section{Outlook}\label{sec:outlook}

The implications of the successful demonstration of \emph{Dichography} extend significantly beyond the physical observations and the overall quality of the two-frame reconstructions discussed in this work. In fact, the practical realization of two-frame CDI from overlapping scattering signals can serve as a strong driver for further improvements in the two-color mode of XFELs, e.g.\ toward higher intensities. Furthermore, it could act as a catalyst for original experimental concepts and designs in several directions that were previously considered nonviable for imaging approaches.

Another modern operational mode of XFELs allows for the generation of light pulses with sub-femtosecond durations \cite{duris2020tunable, maroju2021complex, nolting2023swiss, yan2024terawatt, franz2024terawatt, guo2024experimental, prat2025attosecond}. The combination of these \emph{attosecond} pulses with diffraction imaging has already demonstrated the ability to access ultrafast ionization dynamics by exploiting transient electronic resonances \cite{kuschel2025}.
Recently, \emph{attosecond} pulses have been combined with the \emph{two-color} mode \cite{guo2024experimental}. Thanks to the short time duration of the pulses, which can outrun the lifetime of excited electronic configurations, \emph{two-color attosecond Dichography} may provide unprecedented insights into such elusive states, \newtext{including the possibility to trace the spatial distribution and temporal evolution of electronic excitations} \cite{ulmer2025,zimmermann2025laser} or plasmonic phenomena such as charge density waves in complex nanostructures, depending on their size, shape and surrounding environment. 

Two-color X-ray \emph{Dichography} can be effectively combined with the use of a third pulse from an optical laser to enable a completely new class of imaging experiments. The laser can be employed to trigger a metastable state, which temporarily opens a specific transition triggered with the XFEL \emph{pump} and tracked by the XFEL \emph{probe}. The two consecutive XFEL pulses can then be employed as a \emph{double probe}. For example, nanoparticles can be superheated by \newtext{an intense optical laser pulse}, leading to fast out-of-equilibrium dynamics, like shock-waves and disintegration of the sample \cite{dold2025melting}. 
\newtext{The XFEL pulses can either interact with the sample before and after, or both following, or even overlapping, laser irradiation.} \emph{Dichographic} reconstructions are thus two snapshots of the evolving system that allow for the precise tracking its kinematics and the dynamics of the molecules.

While  the current performance of \emph{two-color} pulses turns out to be challenging for \emph{Dichography}, recent works have demonstrated the capability of partially separating individual photons between the two pulses directly in the \newtext{recorded signal \cite{hecht2025two}.} \newtext{This is achieved by exploiting the sensitivity of the detector to the photon's energy combined with advanced analysis techniques, and can be employed only in the regions of the scattering signal where photon events are sufficiently \emph{rare} \cite{hecht2025two}, typically at high scattering angles.
Integrating this additional information into the algorithmic workflow of \emph{Dichography} is currently under development. Enhancements in the quality and reliability of the reconstructions are foreseen and they will be the subject of future work.}

Improvements of XFELs in \emph{two-color} mode, in terms of pulse brightness, are also expected in the next years, thanks to the developments of XFEL technology encouraged by the increasing interest of the scientific community \cite{vicario2021two}. However, even with the current performance, the brightness of the scattering signal can be boosted in different ways. The xenon-doped helium nanodroplets analyzed in this work have an intrinsically low scattering cross section \cite{marr1976absolute}: the highly-scattering xenon represents only a small fraction of the whole system, thus contributing only marginally to the scattered field. An overall diffraction signal even orders of magnitude higher is expected for systems with significantly higher cross sections, e.g. heavier elements. 
Furthermore, markedly larger scattering cross sections even for lighter materials can be attained at lower photon energies, from a few tens to few hundreds of eV \cite{allaria2015,rossbach2019}. This energy regime in-between the extreme ultraviolet and soft X-ray is also characterized by the presence of electronic resonances for different elements  \cite{henke1993x}, like the biologically relevant carbon and oxygen, where ultrafast electron dynamics can be triggered \cite{kuschel2025} and traced thanks to the strong variations in the scattering cross section \cite{rupp2020imaging}.

A further experimental domain that can benefit from the imaging capabilities of \emph{Dichography} is the study of physical systems sensitive to light polarization. Indeed, several XFEL light sources can partially \cite{lutman2016polarization} or fully \cite{kittel2024demonstration} tune and control the polarization properties of the light in both the XUV \cite{allaria2014control} and X-ray spectra \cite{yan2020polarization}, even with ultrashort light pulses \cite{marotzke2025first}.
Moreover, \emph{two-color} pulses with controllable time-delays and independent polarizations have been recently demonstrated \cite{prat2022widely}. \emph{Dichography} can be combined with these experimental capabilities to study the spatial distribution and time evolution of \emph{dichroic} systems, i.e., systems whose scattering response is sensitive to light polarization. Among these, magnetic domains \cite{stohr1995x}, molecular chirality \cite{cireasa2015probing}, and orbital ordering in crystals \cite{wilkins2003direct} are highly active research topics.

While the difference in wavelength between the two pulses is effective for a better spatial separation of the signal on the scattering detector, it is not a fundamental requirement for \emph{Dichography} \cite{millane2015phase}. In particular, if significant structural modifications are expected to take place in the sample between the two XFEL pulses \cite{sauppe2024double}, the same photon energy can be effectively employed. This feature would allow the whole set of XFEL undulators to be tuned to the same wavelength, and thus the full peak power to be reached. However, the development of methods to produce pairs of electron bunches with sub-picosecond controllable time delay is required to gain full advantage from such an experimental scheme.

\newtext{
\emph{Dichography} can be in principle extended to 3D imaging, for which the imaging problem is more constrained, i.e., \emph{easier} to solve \cite{elser2003phase}. Furthermore, it can be extended to the reconstruction of more than two \emph{frames} \cite{millane2015phase}, especially for sample properties characterized by a high constraint ratio \cite{elser2008reconstruction} (like doped helium nanodroplets, see Sec.~\ref{sec:constrratioDCDI}), exploiting future, more advanced, multicolor capabilities of modern XFELs, like the possibility of delivering more than two different photon energies in the interaction region.}

\emph{Dichography} represents a major step forward in harnessing the imaging capabilities of \emph{two-color} light pulses emerging at XFEL facilities\newtext{, enabling a new class of time-resolved imaging experiments}. Indeed, the two \emph{views} on the sample at different points in time embody, in every respect, one of the original promises of X-ray Free Electron Lasers: recording ultrafast movies of nanomatter.


\subsection*{Acknowledgments}
\vspace{-0.9em}

The research leading to these results was supported by SNSF project 10003697 ``Visualizing coherent and collective electron dynamics in nanoscale matter'' and the COST Action CA21101 ``Confined Molecular Systems: From a New Generation of Materials to the Stars (COSY)''. 
We acknowledge the Paul Scherrer Institute, Villigen, Switzerland for provision of free-electron laser beam time at the Maloja instrument of the SwissFEL ATHOS branch and thank the SwissFEL staff for their assistance.
We acknowledge the European XFEL in \mbox{Schenefeld}, Germany, for the provision of X-ray free-electron laser beam time at the SQS instrument and thank the EuXFEL staff for their assistance. 
TF and CP acknowledge funding by the Deutsche Forschungsgemeinschaft within CRC 1477 "Light-Matter Interactions at Interfaces" (project number 441234705). 
We thank the IT Services Group (ISG) of the Department of Physics at ETH Zurich for the excellent support and management of the computing hardware employed for this research.\\
\newtext{At the time of the original submission, the authors were not aware of the theoretical and algorithmic work of Millane and Chen \cite{millane2015phase}. We thank Prof. Richard Kirian (Arizona State University, USA) for bringing this work to our attention. We are especially grateful to Prof. Rick Millane (University of Canterbury, New Zealand) for the feedback and suggestions on the original version of the manuscript, which significantly contributed to the improved quality of the theoretical arguments presented in this revised version.}

\vspace{-1em}

\subsection*{Author contributions statement}
\vspace{-0.9em}

LH, BB, TB, SDe, ADF, SDo, TF, RH, SK, BK, AØL, BL, SM, TM, MMe, CP, TP, BS, KSi, FSt, RMPT, PT, SU, YO, MMu, DR and AC contributed in the planning and conduction of the experiment at the SQS instrument of the European XFEL (ID 2857). The experiment was led by MMu and YO.
LH, AAH, JB, CB, SDo, TF, FG, CG, GJ, MJ, GK, KK, CP, SR, AS, MS, FSc, KSc, ZS, PT, CFU, VW, XX, MY, OY, NY, HZ, BvI, DR and AC contributed in the planning and conduction of the experiment at the Maloja endstation of SwissFEL (ID 20222035).  The experiment was led by BvI and DR. SD, JB and SR curated the sample delivery, whose performance allowed for the systematic recording of double-hits.
AC developed the original algorithmic approach to \emph{Dichography} and its numerical implementation. AC and DR analyzed the performance of the method on simulated data and ``double-hit'' patterns from the SwissFEL experiment. LH pre-processed the two-color data from the European XFEL experiment, took care of the data selection and developed and implemented the size and shape determination of the helium nanodroplets. AC and LH adapted the DCDI method for \emph{Dichography} and conducted tests on simulated and experimental data. AC, LH and DR contributed in the writing of the manuscript, with input and feedback from all coauthors.

\vspace{-1em}

\subsection*{Data and code availability}
\vspace{-0.9em}

The imaging software employed to produce the results presented here is hosted at \url{https://gitlab.ethz.ch/nux/equinox}, along with examples on simulated and experimental data.
Data recorded for the experiment at the European XFEL are available at \url{doi:10.22003/XFEL.EU-DATA-002857-00}.


\section{Methods}

\subsection{Algorithmic approach to Dichography}\label{sec:dichoalg}

\subsubsection{Iterative phase retrieval}\label{sec:IPR}

\begin{figure*}
\includegraphics[width=0.85\textwidth]{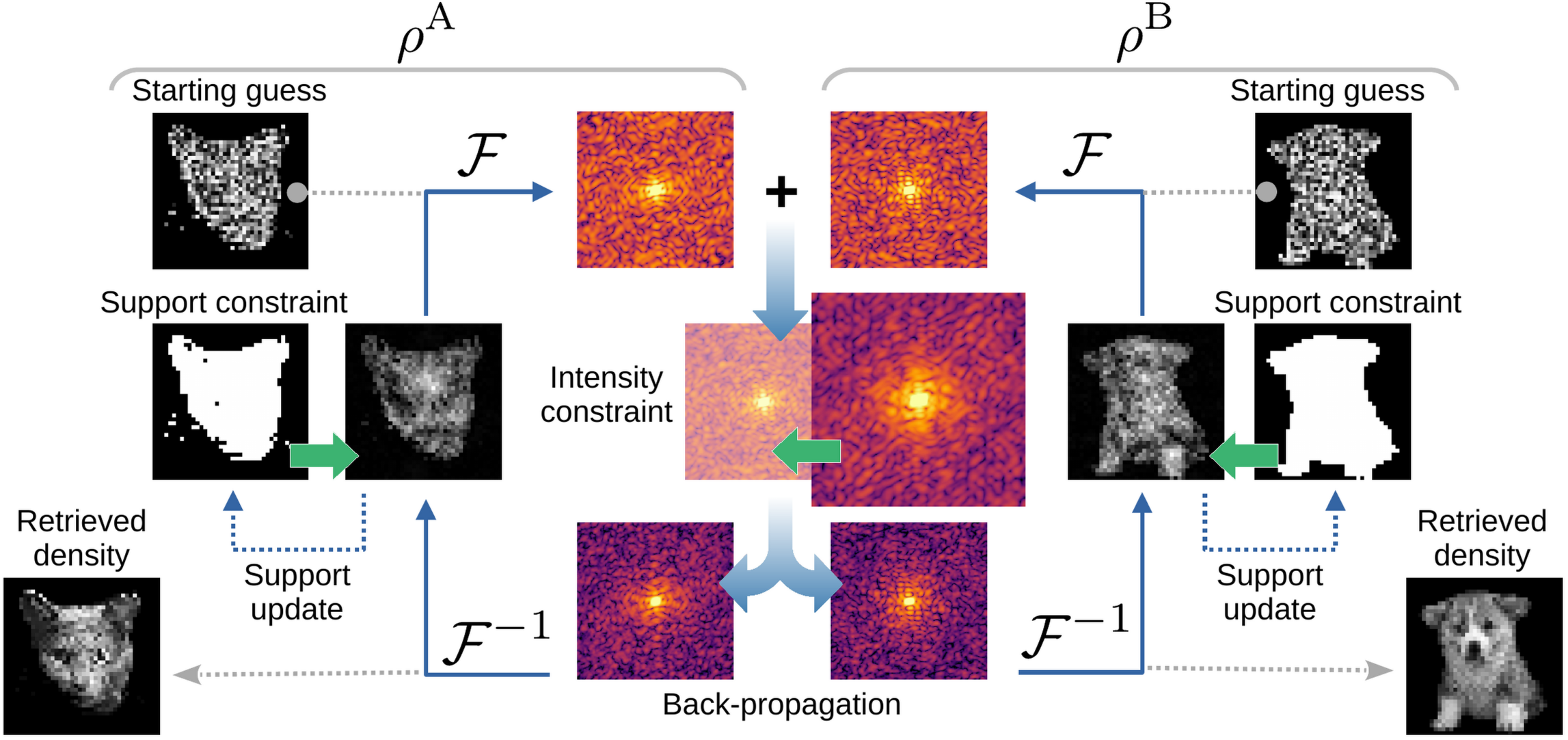}
\caption{Scheme of the iterative phase retrieval algorithm for \emph{Dichography}. The approach corresponds, in most steps, to the execution of two independent reconstruction procedures, $\one$ and $\two$, using conventional iterative phase retrieval algorithms. The two densities, $\den^\one$ and $\den^\two$, are randomly initialized. Two constraints -- the support function in real space and the experimental intensities in Fourier space -- are cyclically enforced. The application of the real-space constraint (shown on the right and left sides of the figure) is independent for $\one$ and $\two$, each with its own support function. This step is thus equivalent to conventional methods for two independent reconstructions. The Fourier intensity constraint, depicted in the center of the figure, is shared between $\one$ and $\two$. Here, the sum of the two Fourier amplitudes is replaced by the experimentally measured values. This information is then back-propagated to the two separate densities, $\den^\one$ and $\den^\two$. The two reconstruction outcomes are returned once the total number of iterations is reached. 
\newtext{The support function is periodically updated and refined, based on the current density estimate, completely independently between the two frames, as discussed in Sec.~\ref{sec:suppconstraint}}
}
\label{fig:flowchart}
\end{figure*}

Coherent Diffraction Imaging (CDI) is an \emph{indirect} imaging method that requires extensive numerical analysis to obtain the sample image from the recorded data.
The scattered field $\psi$ encodes the complete information about the spatial density of the sample $\rho$ as they are mathematically linked by a Fourier Transform (FT) operation (under certain approximations \cite{colombo2023imaging}). It is therefore possible to reconstruct $\rho$ from the full knowledge of the field, i.e., $\rho \propto \mathcal{F}^{-1}\left[ \psi \right]$ where $\mathcal{F}^{-1}$ indicates the inverse FT operation.
However, the experimentally recorded image, $I$, provides information only on the field’s amplitude, i.e., $I \propto \left| \psi \right|^2$, while the field’s phase is lost. Recovering this lost phase corresponds to reconstructing the sample density $\rho$. This so-called \emph{phase retrieval problem} \cite{millane1990phase} lies at the core of CDI analysis and must be solved by suitable numerical approaches.

Conventional algorithms for phase retrieval, such as the Hybrid Input-Output (HIO) and Error Reduction (ER) methods \cite{fienup1982phase}, are characterized by an \emph{iterative} scheme, as they cyclically impose two constraints \cite{marchesini2007invited,elser2003phase, millane2013iterative}. On the one hand, the \emph{oversampling} condition is enforced by setting to zero the density values outside the spatial extension of the sample. This is achieved by defining a \emph{support} function, a binary-valued matrix whose entries are set to $1$ where non-zero density of the sample is expected, and to $0$ where zero scattering is enforced. On the other hand, the intensity constraint is applied in reciprocal space, where the experimentally measured intensities replace the Fourier amplitudes of the current density estimate. This iterative approach is retained for \emph{Dichography} and lies at the core of the numerical retrieval of the two frames, $\rho^\one$ and $\rho^\two$.

Fig.~\ref{fig:flowchart} shows a flowchart of the iterative phase retrieval algorithm adapted to \emph{Dichography}.
It comprises two reconstruction workflows, one for each \emph{frame}, visually separated on the left and right sides of Fig.~\ref{fig:flowchart}. Each workflow follows the typical scheme of iterative projections algorithms \cite{fienup1982phase,marchesini2007invited, elser2003phase}. Both densities, $\rho^\one$ and $\rho^\two$, are independently initialized, typically with random values, and each is constrained to its own support function in real space, $\sup^\one$ and $\sup^\two$, respectively. However, unlike conventional CDI, the two procedures are entangled when constraining the experimental intensities in Fourier space. This step is shown at the center of Fig.~\ref{fig:flowchart}.

For each coordinate $i,j$ in Fourier space, the sum of the squared amplitude of the scattered fields $\psi^\one$ and $\psi^\two$ is constrained to the acquired intensities $I$, i.e.:
\begin{equation}\label{eq:intconstraint}
\left| \psi^\one_{ij} \right|^2  + \left| \psi^\two_{ij} \right|^2 \stackrel{!}{=}  I_{ij} .
\end{equation}
The way in which this is numerically enforced and propagated to the two individual reconstructions is discussed in the next subsection.

\subsubsection{Constraining experimental data}\label{sec:intconstraint}

In \emph{Dichography}, the experimentally acquired information on the sum of the two scattering signals has to be constrained, i.e., propagated to the two frames of the reconstruction, to make the two estimates of the field, $\psi^\one$ and $\psi^\two$, satisfy the condition in Eq. \eqref{eq:intconstraint}. \newtext{Its operation is equivalent to the one proposed in Ref.~\cite{millane2015phase} for the specific case of two incoherent contributions to the scattering.}

\begin{figure}
\includegraphics[width=\columnwidth]{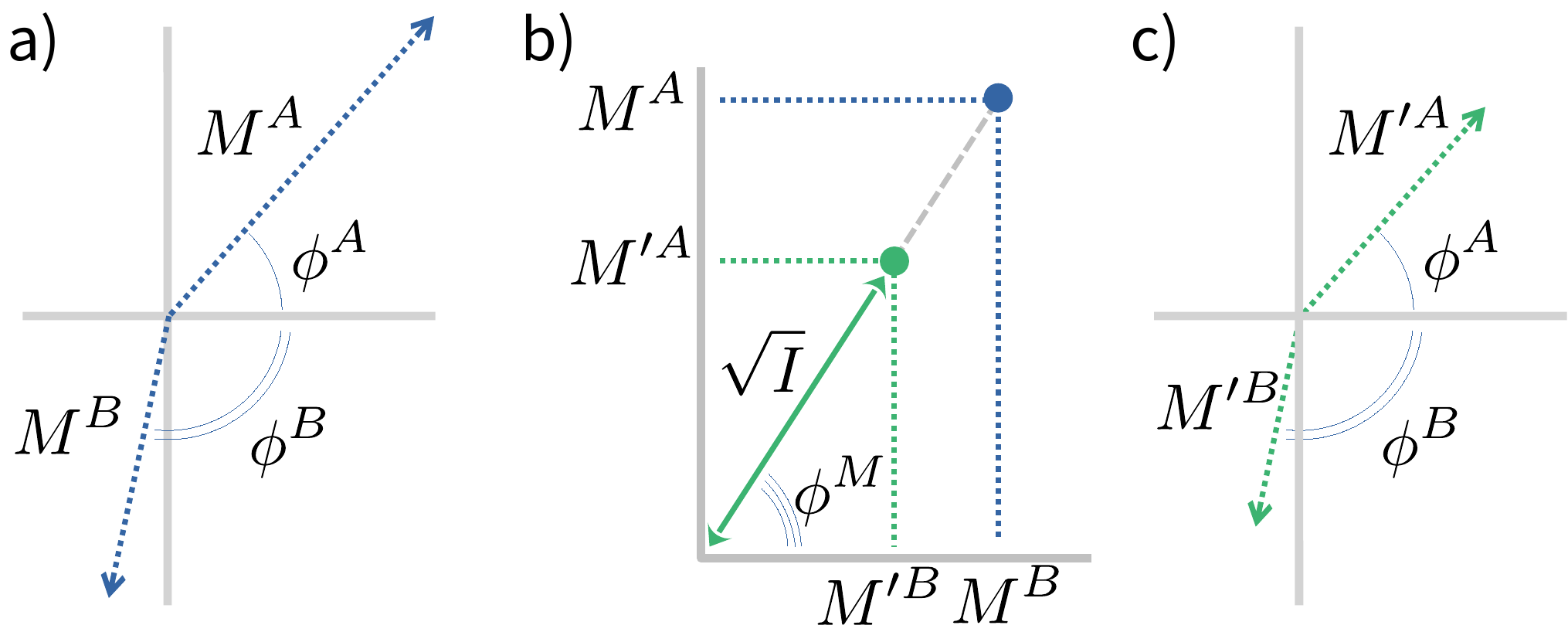}
\caption{Graphical representation of the action of the \emph{intensity projector} as implemented for \emph{Dichography}. In a), the amplitude $M$ and phase $\phi$ of the two reconstructed fields $\psi_{ij}^\one$ and $\psi_{ij}^\two$ are represented in the complex plane. In b), the two amplitudes $M^\one$ and $M^\two$ can be interpreted as the components of a two-dimensional vector, whose norm is rescaled to match $\sqrt{I}$, where $I$ is the acquired experimental data. In c), the rescaled amplitudes $M'^\one$ and $M'^\two$ replace the original ones, while the phases $\phi^\one$ and $\phi^\two$ are retained. This representation of the intensity constraint makes it clear that, for each value of the experimental data $I_{ij}$, three phases must be retrieved, $\phi^\one$, $\phi^\two$ and $\phi^M$, in contrast with a single one in conventional CDI.}
\label{fig:intscheme}
\end{figure}

An intuitive description of the operation is reported in Fig.~\ref{fig:intscheme}. For a given pixel coordinate $i,j$, the values of the scattered fields $\psi_{ij}^\one$ and $\psi_{ij}^\two$ can be represented in the complex plane, as in Fig.~\ref{fig:intscheme}a, in polar coordinates. The two moduli, $M^\one$ and $M^\two$, can then be considered as the two components of a two-dimensional vector, as shown in Fig.~\ref{fig:intscheme}b. The condition in Eq. \eqref{eq:intconstraint} can be achieved by a \emph{renormalization} of the vector $(M^\one, M^\two)$ to obtain a new vector $(M'^\one, M'^\two)$ whose norm corresponds to the square root of the experimental data, $\sqrt{I}$, while its phase $\phi^M$ is preserved. The rescaled moduli $M'^\one$ and $M'^\two$ replace the original ones in the two field estimates, as shown in Fig.~\ref{fig:intscheme}c, while the field phases $\phi^\one$ and $\phi^\two$ are retained. Once the operation is repeated for all entries in the matrix, the experimental constraint is fully \emph{back-propagated} to the two frames of the reconstruction. 

\newtext{
Unlike conventional CDI -- in which only a single phase needs to be recovered for each field amplitude -- the \emph{phase retrieval problem} in \emph{Dichography} is three-fold. Indeed, it requires the recovery of three phase values: $\phi^\one$ and $\phi^\two$ for the two scattered fields, as well as the relative fields amplitude encoded in the ``fictitious'' phase $\phi^M$. However, this observation doesn't imply that the \emph{dichographic} problem is three times harder, thanks to the action of the support constraints \cite{millane2015phase} as further discussed in Sec.~\ref{sec:constrratio}.}
A more detailed presentation of the mathematical operations involved is provided in Sec.~S3 of the Supplemental Material~\cite{supplemental}\newtext{, and thoroughly discussed and extended to an arbitrary number of incoherent scattering contributions in Ref.~\cite{millane2015phase}.}

The most common iterative phase retrieval algorithms share the same way of constraining Fourier intensities, while they differ in the application of the support constraint. The modified intensity constraint described in Fig.~\ref{fig:intscheme} can thus be directly implemented in all conventional iterative phase retrieval algorithms \cite{marchesini2007invited,fienup1982phase, elser2003phase}\newtext{, as shown in the case of the Difference Map \cite{millane2015phase}}. The behavior of the algorithms, however, differs from their conventional CDI counterparts due to the different mathematical nature of the underlying optimization problem (as discussed in Sec.~S4 of the Supplemental Material~\cite{supplemental}).

Conceptually similar implementations of this intensity constraint have been developed and employed to deal with broadband diffraction data, in cases where the spectral information is known \cite{abbey2011lensless, huijts2020broadband}, or even where the spectrum of the scattered light is retrieved along the reconstruction process \cite{chen2024ultra}. These approaches have opened new possibilities for CDI with broadband or polychromatic pulses, like those delivered by High-Harmonic Generation sources \cite{rupp2017coherent}. Crucially, unlike those methods, \emph{Dichography} doesn't enforce any structural connection between the contributions to the recorded scattering signals \newtext{and doesn't involve any \emph{a priori} knowledge on the sample's shape}. This allows capturing structural changes in the samples, as discussed in the following section.

\subsubsection{Real-space constraint}\label{sec:suppconstraint}
The application of the real-space constraint, i.e., the support function, is performed independently on the two frames, as if they were two separate reconstruction procedures using iterative algorithms for conventional single-particle CDI. This means that each of the two densities, $\rho^\one$ and $\rho^\two$, is constrained to its own independent support function, $\sup^\one$ and $\sup^\two$, respectively, as shown in Fig.~\ref{fig:flowchart}. The way the support constraint is enforced depends on the specific iterative algorithm employed \cite{marchesini2007invited,elser2003phase} and retains its form in the adaptation for \emph{Dichography}.

A similar consideration applies to the retrieval of the support function itself. A fundamental challenge in single-particle CDI is the definition of the support function, which is, in most cases, not known \emph{a priori}. The \emph{Shrink-wrap} algorithm (SW) \cite{marchesini2003x} has proven to be a fundamental tool for retrieving the correct support function without prior knowledge of the sample. The SW algorithm refines an initially loose support estimate during the reconstruction process, as described in Ref. \cite{marchesini2003x}. Here, the SW algorithm is employed in its original form to independently update and retrieve the support functions of the two densities, $\rho^\one$ and $\rho^\two$. All \emph{Dichography} results presented in this work used the SW algorithm to determine the samples' spatial extent, and no \emph{a priori} information about the sample geometry \newtext{(or about the structure of xenon doping for DCDI reconstructions)} was employed.

\subsection{Further considerations on the dichographic problem}\label{sec:further}


{
\color{\newtextcolor}
\subsubsection{Constraint requirements for Dichography}\label{sec:constrratio}

In this section we mathematically show that, thanks to their structural properties, the samples reconstructed in this work imprint enough information in the overlapping diffraction patterns to allow for their reconstruction.

In particular, a necessary (yet not sufficient) requirement for the existence of a unique solution to the \emph{dichographic} problem is that the amount of independent information contained in the diffraction data is not smaller than the number of unknowns to retrieve. An effective way to evaluate the ratio between these two quantities is the \emph{constraint ratio} \cite{elser2008reconstruction}. Initially introduced in the context of conventional CDI, it has been further extended to the case of the incoherent sum of $N$ diffraction signals \cite{millane2015phase}. In the specific case of \emph{Dichography}, i.e., $N=2$, the constraint ratio $\Omega$ can be expressed as:

\begin{equation}\label{eq:constrratio}
\Omega = \frac{1}{2} \frac{ \left| \mathcal{A}^\one \cup \mathcal{A}^\two \right| }{ \left| \sup^\one \right|   + \left| \sup^\two \right| }
\end{equation}
where $\sup^\one$ and $\sup^\two$ denote the support functions of the densities $\rho^\one$ and $\rho^\two$, while $\mathcal{A}^\one$ and $\mathcal{A}^\two$ are the supports of the autocorrelation functions of the densities. The cardinality operation, e.g. $\left| \sup \right|$, indicates the number of coordinates belonging to the support functions, i.e., their number of pixels in two dimensions.
A value of $\Omega > 1$ means that the diffraction image contains sufficient information to constrain the density retrieval. Through Eq.~\eqref{eq:constrratio} it is possible to show that the \emph{dichographic} problem is at most a factor 2 less constrained than the conventional CDI case \cite{millane2015phase}.

The value of $\Omega$ can be directly calculated via Eq.~\eqref{eq:constrratio} for the \emph{double-hit} reconstructions in Sec.~\ref{subsec:SwissFEL}, all fulfilling the necessary condition $\Omega>1$. 
The values for Fig.~\ref{fig:SFEL_examples}a, Fig.~\ref{fig:SFEL_examples}b and Fig.~\ref{fig:SFEL_examples}c are $\Omega=1.22$, $\Omega=1.32$ and $\Omega=1.31$, respectively. While similar, they arise from different geometrical features of the samples \cite{millane2015phase}. For example, the similarity in size of the support functions in Fig.~\ref{fig:SFEL_examples}a (which tends to reduce $\Omega$) is compensated for by the distinct overall shapes of the two objects. The $\Omega$ value is similar for Fig.~\ref{fig:SFEL_examples}c, but due to a different reason. Indeed, here, the two support functions are symmetric, reducing the extension of the individual autocorrelation functions and, thus the information content \cite{millane2015phase}. A higher value of $\Omega=1.67$ is calculated for Fig.~\ref{fig:SFEL_examples}d. In fact, the support functions are very different in size and strongly asymmetric for the larger object, increasing the information carried by the autocorrelation.
Among the \emph{double-hit} reconstructions, the one in Fig.~\ref{fig:SFEL_single} is characterized by the smallest value with $\Omega=1.12$. The two support functions are similarly extended and have center-symmetric shapes, rendering the problem less constrained and the reconstruction process more challenging.

The calculation of the constraint ratio for the two-color reconstructions in Fig.~\ref{fig:SQS_results} is, however, not straightforward. A direct application of Eq.~\eqref{eq:constrratio}, considering only the extension of the autocorrelation and the support of the dopants, provides values of $\Omega=6.49$ and $\Omega=12.65$ for the reconstructions of the patterns in Fig.~\ref{fig:SQS_results}a and Fig.~\ref{fig:SQS_results}d, respectively. While those values are particularly high, due to the sparsity of the xenon agglomerates in the droplet, they represent an underestimate of the actual constraint ratio, as they do not take into account the additional information provided by the interference between the xenon density $\rho_\chi$ and the droplet density $\rho_\bigcirc$, the latter being known \emph{a priori} and constrained in the reconstruction. 
In the following section, we introduce a more rigorous method to calculate the constraint ratio for the DCDI approach~\cite{tanyag2015communication}.

\subsubsection{Constraint ratio for Droplet Coherent Diffraction Imaging}\label{sec:constrratioDCDI}

For a conventional DCDI reconstruction~\cite{tanyag2015communication}, both the droplet profile $\rho_\bigcirc$, known \emph{a priori}, and the dopant structure $\rho_\chi$ contribute to the total scattering density $\rho =  \rho_\bigcirc + \rho_\chi$. Their contribution to the autocorrelation function can thus be separated as:

\begin{equation}\label{eq:autocorrDCDI}
\begin{split}
A\left[\rho \right] &= \rho \star \rho \\
&= \rho_\bigcirc \star \rho_\bigcirc +  \rho_\chi \star \rho_\chi + \rho_\chi \star \rho_\bigcirc + \rho_\bigcirc \star \rho_\chi
\end{split}
\end{equation}

Among all contributions to the autocorrelation function, the cross-correlation terms $\rho_\chi \star \rho_\bigcirc$ and $\rho_\bigcirc \star \rho_\chi$ (i.e., the interference terms between the two densities $\rho_\bigcirc$ and $\rho_\chi$) and the dopant autocorrelation $\rho_\chi \star \rho_\chi$ carry information on $\rho_\chi$.

The coordinates at which the autocorrelation function effectively provides information on $\rho_\chi$ are thus those identified by the autocorrelation support:
\begin{equation}\label{eq:autocorrsupportDCDI}
\mathcal{A}_{\bigcirc\chi} = (A\left[\rho \right] - A\left[\rho_\bigcirc \right]) > 0
\end{equation}

As the information to retrieve corresponds to the support function $\sup_\chi$ of the dopant density $\rho_\chi$, the constraint ratio for conventional DCDI follows as:
\begin{equation}\label{eq:constrratioDCDI}
\Omega_{\bigcirc\chi} =  \frac{ \left| \mathcal{A}_{\bigcirc\chi} \right|}{ 2\, \left| \sup_\chi\right|}
\end{equation}

It is worth noting that the presence of the droplet increases the information content of the diffraction through the interference terms. Thanks to this \emph{holographic} information, the imaging problem is significantly more constrained, which readily explains the exceptional convergence properties of the DCDI algorithm observed in previous works \cite{tanyag2015communication}.

It is now possible to extend the constraint ratio $\Omega_{\bigcirc\chi}$ to the \emph{Dichography} problem, following the arguments discussed in Ref.~\cite{millane2015phase}. For $N$ scattering contributions, the constraint ratio is evaluated as:
\begin{equation}\label{eq:constrratioDCDI_n}
\Omega_{\bigcirc\chi} = \frac{ \left| \bigcup\limits_{n=1}^N \mathcal{A}_{\bigcirc\chi}^n\right|}{ 2\, \sum\limits_{n=1}^N \left| \sup_\chi^n \right|}
\end{equation}
which reduces for \emph{Dichography} ($N=2$) to:
\begin{equation}\label{eq:constrratioDCDI_dicho}
\Omega_{\bigcirc\chi} = \frac{1}{2} \frac{ \left| \mathcal{A}_{\bigcirc\chi}^\one \bigcup \mathcal{A}_{\bigcirc\chi}^\two \right|}{  \left| \sup_\chi^\one \right| + \left| \sup_\chi^\two \right|}
\end{equation}

The constraint ratio adapted to DCDI in Eq.~\eqref{eq:constrratioDCDI_dicho}, which takes into account the additional information contained in the interference terms, yields values of $\Omega_{\bigcirc\chi}=9.57$ and $\Omega_{\bigcirc\chi}=19.73$ for the reconstructions from Fig.~\ref{fig:SQS_results}a and Fig.~\ref{fig:SQS_results}d, respectively. As expected, those values are higher than those obtained by a direct use of the original formula in Eq.~\eqref{eq:constrratio}.

It is worth noting that the constraint ratio $\Omega$ derives purely from geometrical properties of the samples, and it does not include an evaluation of the quality of the information contained in the diffraction data, in particular concerning the amount of noise due to photon statistics. The inclusion of noise and missing data in the evaluation of the constraint ratio is currently an open problem even in conventional CDI, and it will be investigated in future works.
Furthermore, $\Omega>1$ represents a necessary, but in general not sufficient, condition for the uniqueness of the solution. 

\subsubsection{Non-existence of a conventional CDI solution}\label{sec:cdisol}

In conventional CDI, the failure of a reconstruction attempt is a typical indicator of a double-hit event. This empirical observation raises the question whether this behavior is systematic. Does a conventional CDI solution exist for the incoherent sum of two diffraction patterns? Here, we formally approach this question via polynomial decomposition (following Ref.~\cite{hayes1982reconstruction}). We conclude that, in general, such a ``fictitious'' conventional solution does not exist.

Let $\rho^\one(\vec{x})$ and $\rho^\two(\vec{x})$ be two distinct discrete objects (where $\vec{x}$ indicates the discrete representation of the spatial coordinates), with finite support in a $d$-dimensional space ($d \ge 2$). We, thus, exclude the trivial case of identical densities, for which there is no difference between coherent and incoherent sum. The case of an identical physical particle imaged with two different colors is, however, still included, as the same particle is differently sampled, thus corresponding to the case of two different densities.

Their Fourier intensities are given by $I^\one = \tilde{\rho}^\one \tilde{\rho}^{A*}$ and $I^\two = \tilde{\rho}^\two \tilde{\rho}^{B*}$.
We seek a single effective compact object $\rho^{\text{eff}}(\vec{x})$ such that:
\begin{equation}
\tilde{\rho}^{\text{eff}}(\vec{q})\tilde{\rho}^{\text{eff}}(\vec{q})^*   \stackrel{?}{=} \tilde{\rho}^\one(\vec{q})\tilde{\rho}^\one(\vec{q})^* + \tilde{\rho}^\two(\vec{q}) \tilde{\rho}^\two(\vec{q})^* \quad \forall \vec{q}.
\end{equation}

The Discrete Fourier Transform (DFT) operation corresponds to the Z-transform evaluated on the unit circle (specifically the unit torus in $d$-dimensions). In particular, for the 2D case, the discrete density function $\rho(\vec{x})$ is mapped to a polynomial $P(\vec{z})$ via:
\begin{equation}
P(z_1, z_2) = \sum_{x_1, x_2 \in \sup} \rho(x_1, x_2) z_1^{-x_1} z_2^{-x_2},
\end{equation}
where $\sup$ is the finite support. The diffraction pattern $I$ of an individual density $\rho$ corresponds to the polynomial product $P(\vec{z}) P^*(\nicefrac{1}{\vec{z}^*})$ evaluated on the unit circle, i.e., at coordinates $z_j = e^{\text{i} k_j}$.

Finding a single object $\rho^{\text{eff}}$ requires finding a polynomial $P^{\text{eff}}$ that factorizes the sum of the two diffraction patterns, i.e.:
\begin{equation}
P^{\text{eff}}(\vec{z}) {P^{\text{eff}}}^*(\nicefrac{1}{\vec{z}^*}) = P^\one (\vec{z}) {P^\one}^*(\nicefrac{1}{\vec{z}^*})  + P^\two (\vec{z}) {P^\two}^*(\nicefrac{1}{\vec{z}^*}) .
\label{eq:poly_sum}
\end{equation}

In practical terms, Eq.~\eqref{eq:poly_sum} states that a CDI solution to the Dichography problem, $\rho^\text{eff}$, exists only if the sum of the two factorizable polynomials is itself a factorizable polynomial of the form $P^{\text{eff}}(\vec{z}) {P^{\text{eff}}}^*(\nicefrac{1}{\vec{z}^*})$.

For the 2D case, and in general for any dimension $d\geq 2$, Hayes proved \cite{hayes1982reconstruction} that the set of factorizable polynomials forms a set of measure zero in the space of all polynomials.
Thus, almost always, the sum $P^\one {P^\one}^* + P^\two {P^\two}^*$ cannot be expressed as $P^{\text{eff}}{P^{\text{eff}}}^*$. This is analogous to the sum of two squared integers, which is almost never a perfect square.
As a consequence, the measured intensity $I^{\text{tot}}$ does not correspond to the Fourier transform of any single compact object. The data violates the fundamental consistency constraints of the standard phase retrieval problem (specifically the compactness of the density), and a conventional CDI solution cannot be found. This aspect is further investigated numerically in Sec. 6 of the Supplemental Material, along with the discussion of pathological cases for which, due to strict symmetry features, a conventional CDI solution indeed exists.

\subsubsection{Uniqueness of the solution}\label{subsec:uniqueness}

A fundamental aspect of optimization problems is the uniqueness of the optimal solution. Are there equivalent solutions to the \emph{dichographic} optimization problem? Specifically, are the two densities $\rho^\one$ and $\rho^\two$ retrieved by Dichography unique? While Hayes' theorem \cite{hayes1982reconstruction} guarantees the uniqueness of phase retrieval for a single compact object in dimensions $d \ge 2$, the incoherent sum of two intensities introduces an additional degree of freedom. And, indeed, it is easy to verify that a \emph{rotation} (or, more generally, a unitary transform \cite{millane2015phase}) of the densities produces the same diffraction pattern. Specifically, it is possible to create a new pair of frames $\rho'^{\one}$ and $\rho'^{\two}$ via the linear operation:
\begin{equation}\label{eq:rotation}
\begin{pmatrix} \rho'^{\one}(\vec{x}) \\ \rho'^{\two}(\vec{x}) \end{pmatrix} = 
\begin{pmatrix} \cos\theta & \sin\theta \\ -\sin\theta & \cos\theta \end{pmatrix} 
\begin{pmatrix} \rho^\one(\vec{x}) \\ \rho^\two(\vec{x}) \end{pmatrix}.
\end{equation}

These two densities fully satisfy the experimental constraint $|\tilde{\rho}'^{\one}|^2 + |\tilde{\rho}'^{\two}|^2 = I^{\text{tot}}$. Here, the range of angles for which the ambiguity is relevant is restricted to $\theta \in \left[0, \pi/2 \right)$. 
The other regions of possible rotations can be easily related to this interval by combining the operation of \emph{frame swapping}, under which Dichography is inherently ambiguous, or the assignment of a global sign to the density values, under which even conventional phase retrieval is ambiguous.

Mathematically, the diffraction data alone cannot distinguish the true densities from a rotation in Eq.~\eqref{eq:rotation}. 
There are, however, physical constraints that break this rotational equivalence, rendering the solution unique (up to the ambiguities that are also affecting conventional phase retrieval).
A key aspect is the role of the support functions in combination with the constraint of a positive density, known as \emph{positivity constraint} \cite{marchesini2007invited}. Given $\sup^\one$ and $\sup^\two$, we can identify three possible cases:

\paragraph{Incompatible supports} In this case $\sup^\one \neq \sup^\two$ and neither support contains the other. Only the rotation $\theta = 0$ renders the reconstructions compatible with their own support function. Any other rotation would require a support function $\sup = \sup^\one \cup \sup^\two$ for both densities. This is further enhanced by the positivity constraint. Indeed, any rotation apart from $\theta = 0$ would produce a reconstruction with negative density values, forbidden by the constraint.
\paragraph{Nested supports} In this case, $\sup^\one \neq \sup^\two$ but $\sup^\one \subset \sup^\two$. Again, the rotational symmetry is broken, but in a ``milder'' way. Indeed, while a rotation would render $\rho'^{\one}$ incompatible with $\sup^\one$, $\rho'^{\two}$ would still be compatible with $\sup^\two$, as $\sup^\two = \sup^\one \cup\, \sup^\two$. In this case, the positivity constraint doesn't necessarily help as there can be rotations that do not yield negative values. The same considerations apply by swapping the frames $\one$ and $\two$.
\paragraph{Identical supports} In this case $\sup^\one = \sup^\two$ and the rotational ambiguity cannot be resolved by the support functions. A unique solution can be identified thanks to the sole positivity constraint under specific conditions \cite{millane2015phase}. 

It is worth noting that, even in the case of two-color pulses hitting a single particle that retains the shape during the two diffraction events, the support functions of the two reconstructed frames have different spatial extensions. Indeed, the two pulses sample the particle at different resolutions, producing two different discrete representations of the reconstructed densities (see Sec.~\ref{subsec:twocolorrecs}). Thus, the incompatibility of the support functions that resolves the \emph{rotational ambiguity} is inherently provided by two-color pulses (i.e., two-color data generally cannot fall into the tricky ``identical supports'' case).

The discussion so far has been based on the \emph{a priori} knowledge of the support functions $\sup^\one$ and $\sup^\two$. In typical cases, those support functions are not known and they are retrieved during the reconstruction process with the \emph{Shrink-wrap} algorithm \cite{marchesini2003x}. In theory, the \emph{Shrink-wrap} algorithm doesn't prevent the possibility to reach two support functions $\sup'^{\one}= \sup'^{\two} = \sup^\one \cup\,\sup^\two$, thus restoring the rotational ambiguity. However, in practical cases, we observe that the \emph{Shrink-wrap} tends to minimize the extension of the two support functions, i.e., tends to achieve the lowest possible spatial extension of the two frames. This action of the \emph{Shrink-wrap} naturally pushes the reconstruction process away from the $\sup^\one \cup \sup^\two$ case, thus resolving the ambiguity in both the ``Incompatible supports'' and ``Nested supports'' cases.

The \emph{rotational} ambiguity is, thus, theoretically resolved by the support constraint. For ``ideal'' data, not affected by noise, the only zero-error solution is the one where $\theta=0$ and the two frames are perfectly separated. Real experimental diffraction patterns are, however, affected by shot noise, which renders even the correct solution not fully compatible with the constraints. This leaves ``room'' for other reconstructions close to the real solution to arise, similarly to what happens with the \emph{twin-image problem} in conventional CDI \cite{guizar2012understanding}. In the case of Dichography, those reconstructions close-by in the error landscape are represented by slight rotations in Eq.~\eqref{eq:rotation}. 
We believe that these small, unresolved rotations are the origin of the \emph{ghosting} artifacts observed in the experimental reconstructions, where a faint ``ghost'' of one frame appears in the other.
}

\subsection{Adaptation of Memetic Phase Retrieval}\label{sec:MPR}

The capability of iterative phase retrieval algorithms to reach the \emph{solution} to the phase retrieval problem -- i.e., to correctly reconstruct the sample density -- is a key factor in CDI. These algorithms are, in fact, the \emph{virtual lenses} of CDI, and the quality and reliability of the retrieved sample image are heavily dependent on their performance. For this reason, significant effort has been devoted to the development of various iterative algorithms \cite{marchesini2007invited}. These algorithmic approaches are well established and are routinely employed in CDI-based experimental techniques at synchrotrons, such as \emph{Bragg-CDI} \cite{favre2010analysis} or \emph{Ptychography} \cite{guizar2021ptychography}.

Single-particle single-shot CDI performed at XFELs, however, comes with additional complications that render the algorithm's task non-trivial. For example, the central region of the diffraction pattern cannot be recorded due to the presence of a hole in the detector, necessary to prevent damage from the transmitted beam. Furthermore, the brightness of the patterns is most of the time sub-optimal and cannot be increased, as the intensity of the single XFEL pulse is already tuned to its maximum performance. These factors significantly impact the performance of iterative algorithms and render CDI at XFELs considerably more challenging.

A recent breakthrough in algorithmic capabilities comes from the development of the Memetic Phase Retrieval (MPR) approach \cite{colombo2017facing,colombo2025spring}. This approach takes advantage of the parallel execution of several reconstruction procedures on the same diffraction pattern. Information between these reconstructions is shared through a genetic algorithm, which augments the capabilities of conventional iterative algorithms and boosts the quality and reliability of the image reconstruction \cite{colombo2025spring}.

Mathematical considerations (see \methods~\ref{sec:constrratio}) as well as systematic tests on simulated data (see Sec.~S1 and Sec.~S4 of the Supplemental Material~\cite{supplemental}) show that \emph{Dichography} reconstructions are more demanding than conventional CDI, and that the performance of conventional iterative algorithms is further impacted. Thus, the MPR scheme has been adapted to \emph{Dichography} resulting in a significant boost in performance, \newtext{especially concerning the retrieval of the support function during the reconstruction process.
}

The MPR adaptation for \emph{Dichography} is called \emph{Equinox}. \emph{Equinox} has been employed to obtain all imaging results presented in this manuscript, as well as those on simulated data discussed in the Supplemental Material, unless specified otherwise. It is implemented as a Python package directly derived from SPRING~\cite{colombo2025spring}, a recently released toolset for image reconstruction in conventional single-particle single-shot CDI. The open-source implementation of \emph{Equinox} and its documentation are accessible at \url{https://gitlab.ethz.ch/nux/equinox}.

\newtext{
\emph{Equinox} retains the conceptual structure of its conventional CDI counterpart, SPRING \cite{colombo2025spring}. The most significant modification, aside from the use of dichographic iterative projectors (see Sec.~\ref{sec:dichoalg}), lies in the initialization of the starting densities. As in SPRING, the individual frames are initialized by populating the real-space matrix with spherical density profiles of random position and size. In \emph{Equinox}, however, the densities of the two frames are additionally renormalized to reflect unequal scattering contributions. The intensity ratio is randomly assigned within a range from 50:50 to 90:10. While these limits are arbitrary and tunable, we empirically observed that this range is sufficiently robust to reconstruct data with diverse intensity ratios without the need for case-specific fine-tuning. The algorithm can successfully retrieve the correct scattering ratio independent of the initial balancing.}

\newtext{
Currently, \emph{Equinox} implements the Error Reduction (ER), Hybrid Input-Output (HIO) \cite{fienup1982phase}, and Relaxed Averaged Alternating Reflections (RAAR) \cite{luke2005relaxed} algorithms. Future work includes the implementation of the Difference Map (DM) algorithm \cite{elser2003phase} adapted for the \emph{dichographic} problem \cite{millane2015phase}, which is expected to further improve the convergence properties of the solver.
}

\subsection{Experimental details}\label{sec:expdet}

\subsubsection{Silver nanoparticles at SwissFEL}\label{subsec:SwissFELExp}

The experiment was performed at the Maloja endstation of SwissFEL~\cite{sun2022ultrafast,maloja}. 
Wet-chemically grown silver nanoparticles were injected into the experimental chamber, kept under vacuum conditions, using an electrospray and a set of aerodynamic lenses~\cite{bielecki2019electrospray,hantke2018rayleigh}. The electrospray disperses the liquid containing the suspended nanoparticles into sub-micron-sized droplets. The nanoparticle concentration is optimized for a single sample per droplet. The droplets flow toward the interaction region along with a buffer gas in a set of aerodynamic lenses, where the liquid evaporates.

The XFEL pulses, with a duration of \SI{50}{\femto\second} and a total energy of \SI{3}{\milli\joule}, are delivered by SwissFEL at a repetition rate of \SI{100}{\hertz}. The pulses are tuned to a photon energy of \SI{1000}{\electronvolt}, corresponding to a wavelength of \SI{1.24}{\nano\meter}, and intercept the nanoparticles in a \SI{3}{\micro\meter} focus. These parameters correspond to a maximum intensity in the focus of approximately $10^{18}\,\unit{\watt\per\centi\meter^2}$.
The scattering signal is recorded by a 4-megapixel \emph{Jungfrau} detector optimized for soft X-ray photon detection~\cite{mozzanica2018jungfrau,hinger2022advancing}. The detector placement allows for the detection of the diffraction signal up to a scattering angle of \ang{13.5}.

The minimum spatial separation in the transverse direction between the two particles of the double-hit, necessary to avoid the recording of the interference term, is calculated to be \SI{4}{\micro\meter}. The data employed for the reconstructions was further rescaled in the post-processing phase, reducing the distance down to \SI{1}{\micro\meter}. Please refer to Sec.~S2 of the Supplemental Material for further discussion. These quantities, and especially the \SI{4}{\micro\meter} value for the raw detector data, have to be considered an upper bound. The minimum distance can be further reduced by the limited temporal and spatial coherence of the FEL pulse, assumed to be fully coherent in those calculations.

\subsubsection{Two-color experiment at the European XFEL}\label{subsec:SQS}

The experiment was performed at the Nano-sized Quantum Systems (NQS) endstation of the Small Quantum Systems (SQS) instrument~\cite{SQS, tschentscher2017photon, mazza2023}, located on the SASE3 branch of the European XFEL~\cite{decking2020mhz}  \newtext{(proposal 2857)}. The XFEL was operated at a repetition rate of \SI{10}{\hertz} with an electron energy of \SI{16}{\giga\electronvolt}.

Two consecutive sets of XFEL undulators were tuned to \SI{992}{\electronvolt} and \SI{1192}{\electronvolt}, respectively. A magnetic chicane located after the first undulator section delayed the arrival of the electron bunch at the second undulator, introducing a time delay of \SI{750}{\femto\second} between the \SI{992}{\electronvolt} pulse and the subsequent \SI{1192}{\electronvolt} pulse. This enables a \emph{pump--probe} scheme in which both X-ray pulses are generated by the XFEL~\cite{serkez2020opportunities}.

The two pulses, each with a nominal duration of \SI{20}{\femto\second}, were focused to a \SI{5}{\micro\meter} spot in the interaction region~\cite{mazza2023,serkez2020opportunities}, where they intercepted isolated superfluid helium nanodroplets doped with xenon atoms. The scattered light from doped helium nanodroplets, generated by both pulses, was recorded by a pnCCD detector \cite{kuster2021} positioned at the back of the interaction region. The geometry allows capturing scattering signals up to an angle of \SI{6.4}{\degree} at the edge of the detector.

In this special XFEL operation mode, the intensity at the focus is significantly reduced because each pulse is produced and amplified by only half of the available undulator sections, which are tuned to different wavelengths~\cite{serkez2020opportunities}. Consequently, the energy per pulse is reduced to \SI{0.7}{\milli\joule}, compared to more than \SI{4}{\milli\joule} in standard single-color operation. Furthermore, the physical separation between the undulator sections means the two collinear pulses are focused at slightly different downstream positions. As a result, the optimal focus for two-color operation lies between these two positions, with an estimated effective focal spot size of \SI{14}{\micro\meter}~\cite{hecht2025two}.
The combined effect of reduced pulse energy and increased focal spot size yields a nominal maximum intensity of approximately $10^{16}\,\unit{\watt\per\centi\meter^2}$, about two orders of magnitude lower than that achieved in single-color mode.
\newtext{
Data acquisition for doped droplets was limited by technical constraints and only approximately 600 hits could be recorded. From these, the final dataset was selected based on total signal brightness and, crucially, the intensity balance between the two pulses, which varies due to the spatial separation of the two foci \cite{hecht2025two}. Insights into the data selection process are reported in Sec. S2 of the SM.
}

\subsection{Retrieving the shape of helium droplets}\label{sec:fitting}

The Droplet-CDI (DCDI) method \cite{tanyag2015communication} constrains the density distribution of the helium droplet during the reconstruction process, such that the \emph{phase retrieval} task only involves reconstructing the density values of the inner xenon doping. The use of the DCDI method made it possible to analyze the two-color diffraction images with \emph{Dichography}, as it greatly reduces the complexity of the imaging problem and makes it significantly more resilient to noise.

The applicability of the DCDI method to the two-color \emph{pump-probe} data relies on two fundamental prerequisites: the ability to determine the droplet size \emph{a priori}, and the structural integrity of the droplet, which must be preserved over the \SI{750}{\femto\second} time delay between the two XFEL pulses.
Both of these conditions have been verified and demonstrated in Ref.~\cite{hecht2025two}, and the main findings are summarized below.

For a spherical particle, the scattered field can be calculated exactly and analytically using Mie theory, which provides solutions to Maxwell’s equations \cite{Mie1908}. The scattering signal can be fully computed with knowledge of the droplet size $D$, the radiation wavelength $\lambda$, the refractive index of the material $n^{\text{He}}(\lambda)$ at the given photon energy, and the energy density of the light field.

For a diffraction signal from the same pure-helium droplet, composed of the incoherent sum of two different colors, Ref. \cite{hecht2025two} shows that it is possible to retrieve the size of the droplet and the intensity of the two pulses by fitting the recorded radial intensity profile as the incoherent sum of two Mie-calculated intensities, one for each color. In this procedure, the droplet size $D$ is assumed constant between the two scattering events, while the wavelengths of the two pulses, $\lambda^\one = \SI{1.25}{\nano\meter}$ and $\lambda^\two = \SI{1.04}{\nano\meter}$, as well as the corresponding refractive indices for helium, $n^{\text{He}}(\lambda^\one)$ and $n^{\text{He}}(\lambda^\two)$ \cite{henke1993x}, are known \emph{a priori}.

Furthermore, the comparison between the fitted scattering profiles and the experimental data reveals full compatibility of the recorded signal with that produced by a hard sphere, even for the signal produced by the late pulse arriving on the sample after \SI{750}{\femto\second}. This compatibility is experimentally verified up to a spatial resolution of \SI{19}{\nano\meter}, corresponding to the momentum transfer for the \SI{1192}{\electronvolt} photon energy at a \SI{1.5}{\degree} scattering angle. This angle has been identified as the maximum at which the signal-to-noise ratio in the recorded data is sufficiently high to provide useful information, as discussed in Ref. \cite{hecht2025two}.

The observation about the integrity of helium droplets can be supported by theoretical considerations. The maximum number of electrons $n_{e,out}$ that can leave the droplet due to the interaction with the first pulse can be calculated via $n_{e,out} = (h\nu - I_p) 4 \pi \epsilon_0 R e^{-2}$ \cite{bostedt2010a}. For a droplet with radius $R = 350\,$\unit{\nano\meter} and photon energy \SI{992}{\electronvolt}, this corresponds to $n_{e,out} = 2.38 \times 10^5$ charges. This value fixes the maximum kinetic energy of a helium ion leaving the droplet and sets an upper bound to the maximum distance reached by the ion in \SI{750}{\femto\second}, estimated as \SI{18.6}{\nano\meter} (see Sec. S8 of the Supplemental Material \cite{supplemental}). This upper bound is already below the \SI{19}{\nano\meter} resolution limit of the experiment.

The physical system is thus compatible with a compact spherical helium nanodroplet for both the \emph{pump} and the \emph{probe} pulses, as supported by both experimental observations and theoretical considerations. Furthermore, it is possible to extract the droplet size in a precise and consistent way by fitting Mie solutions to Maxwell’s equations for a signal produced by two different photon energies. For a thorough investigation of the physical system and the fitting procedure, please refer to Ref. \cite{hecht2025two}. This provides the \emph{a priori} information necessary for the use of the DCDI method during the imaging process with \emph{Dichography}.

As discussed in Sec.~\ref{sec:suppconstraint}, the iterative approach to \emph{Dichography} retains the real-space operations of the conventional iterative phase retrieval algorithms for CDI. The DCDI method is, in practice, an additional real-space constraint and can be directly implemented in the algorithm workflow (see Fig.~\ref{fig:flowchart}). The only required additional step is to properly adapt the droplet size of the density profile to the corresponding spatial resolution for the two frames, $\rho^\one$ and $\rho^\two$.

\subsection{Representation of two-color reconstructions}\label{subsec:twocolorrecs}

\begin{figure}
\includegraphics[width=\columnwidth]{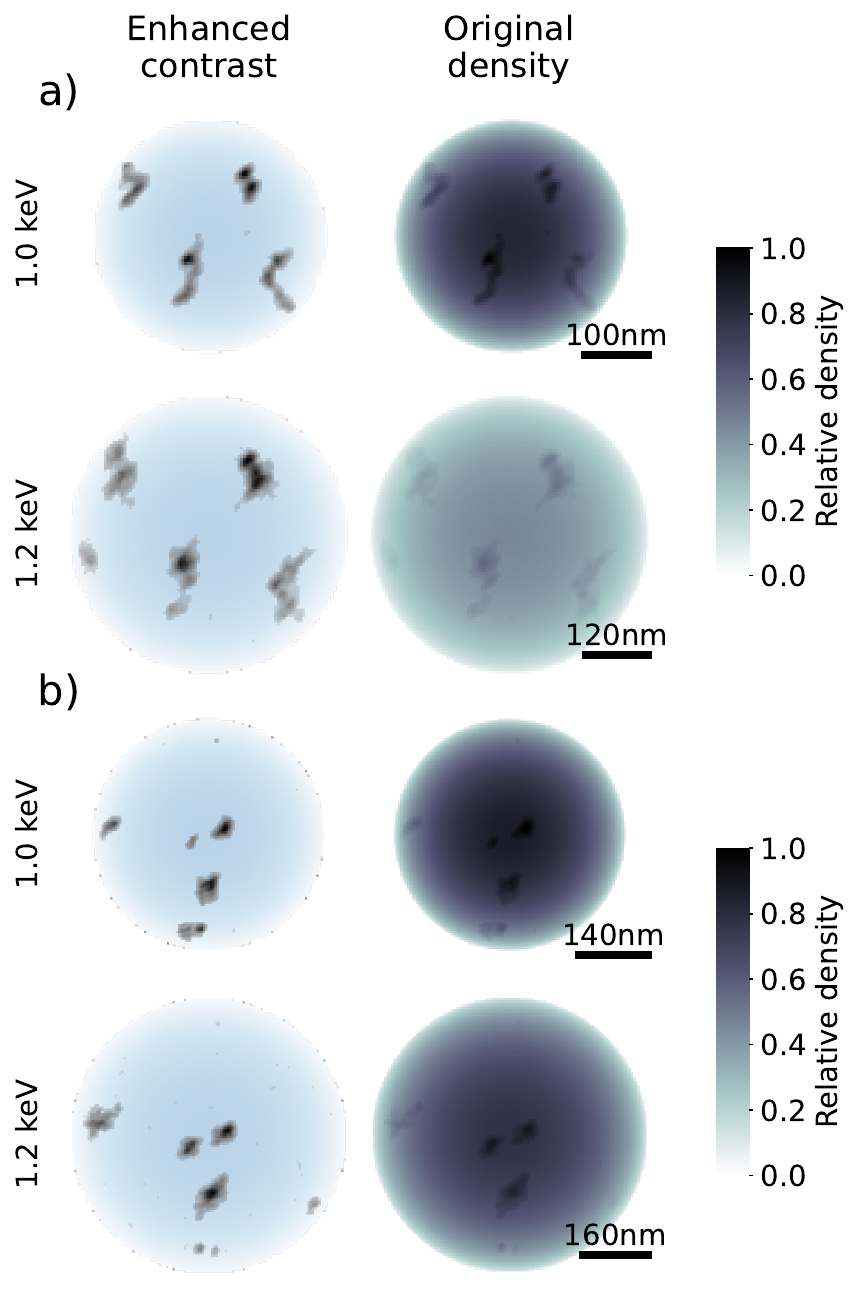}
\caption{\newtext{Two-color reconstructions of the xenon-doped helium nanodroplets shown in Fig.~\ref{fig:SQS_results}, here reported with the original sampling. The figure presents the same data as shown in Fig.~\ref{fig:SQS_results}. Each two-frame reconstruction is reported twice. On the left, the densities are reported with enhanced contrast, similarly to Fig. \ref{fig:SQS_results}, to highlight the inner xenon structures. On the right, the original scattering density is reported. In this second case, the color scale is shared between the two frames to also report the imbalance between the scattering contributions.}}
\label{fig:SQS_results_scaled}
\end{figure}

The spatial extension of a single pixel in the reconstruction corresponds to the \emph{half-period} resolution \cite{colombo2023imaging}. In the \emph{small-angle} approximation \cite{kirian2020imaging}, the spatial extension of a single reconstruction pixel $\Delta$ can be calculated as:
\begin{equation}\label{eq:pixelsize}
\Delta = \frac{\lambda}{2 \theta_\text{max}}
\end{equation}
where $\lambda$ is the radiation wavelength and $\theta_\text{max}$ is the angle measured from the center of the detector to its side edge. For a two-color reconstruction with \emph{Dichography}, the two frames correspond to two views of the same sample obtained with different colors. As such, the corresponding pixel size $\Delta$, i.e., the sampling in real space, differs. In particular, the values for the \SI{1.0}{\kilo\electronvolt} ($\lambda^\one = \SI{1.25}{\nano\meter}$) pulse and the \SI{1.2}{\kilo\electronvolt} ($\lambda^\two = \SI{1.04}{\nano\meter}$) pulse can be calculated via Eq.~\eqref{eq:pixelsize} as $\Delta^\text{\pump} = \SI{5.57}{\nano\meter}$ and $\Delta^\text{\probe} = \SI{4.64}{\nano\meter}$.

As a consequence, when reconstructing a sample of size $D$, the \pump reconstruction will have an extent of $\sfrac{D}{\Delta^\text{\pump}}$ pixels, while the \probe reconstruction will extend over $\sfrac{D}{\Delta^\text{\probe}}$ pixels. 
To better highlight this aspect, we report \newtext{in Fig.~\ref{fig:SQS_results_scaled} the same two-color reconstructions as shown in Fig.~\ref{fig:SQS_results}}, where the two retrieved frames are not scaled to match the spatial extension; i.e., they are cropped from the reconstructed matrices $\rho^\one$ and $\rho^\two$ with the same pixel dimensions. On the one hand, this representation helps the reader visualize the lower resolution of the \pump pulse. On the other hand, it highlights that the appearance of two similar structures in the two frames cannot derive from a \emph{ghosting} effect. In fact, the arising of a reconstruction feature in the wrong frame wouldn't match the proper spatial resolution, since the two structures differ in pixel extent.
\newtext{
In addition, Fig.~\ref{fig:SQS_results_scaled} reports the two scattering densities without the enhancement of the contrast for the xenon structures and with a common color scale between the two frames. The reconstructed densities are reported with the correct scaling of the helium droplet $\rho_\bigcirc$ and the xenon doping $\rho_\chi$, i.e.,  $\rho = \rho_\bigcirc + \rho_\chi$ . This allows the reader to better appreciate the relatively low contrast of the xenon filaments within the droplets, and thus better evaluate the noise visible in the reconstructions shown in enhanced contrast in Fig.~\ref{fig:SQS_results}. Furthermore, it is worth noting that the scattering density $\rho$, outcome of the reconstruction, is proportional to the actual electronic density $\rho_e$ and the field amplitude $\sqrt{I}$, i.e., $\rho \propto \sqrt{I} \cdot \rho_e$. The visibly fainter \probe reconstruction in Fig.~\ref{fig:SQS_results_scaled}a comes indeed from the lower intensity contribution of the \probe pulse in the scattering, which corresponds to a weaker \probe field in the specific position of the imaged droplet within the two focal regions.}



\bibliography{bibliography}


\end{document}